

%

%
%
%

\input phyzzx
\hsize=6.6in

\nopubblock

\rightline{ YITP-96-9 }
\rightline{February 29, 1996}

\titlepage

\title{\bf 
The spectral representation of the spacetime structure: 
The `distance' between universes with different topologies}

\author{Masafumi Seriu}
\address{Inter-University Centre for Astronomy and Astrophysics\break
        Post Bag 4, Ganeshkhind, Pune 411007, India\break}
\andaddress{
  Yukawa Institute for Theoretical Physics \qquad \quad\break
   Kyoto University, Kyoto 606, Japan
   \footnote\star{\rm Present address.}
   }

\vskip 9cm
To appear in {\sl Physical Review D}.
\endpage

\abstract{
We investigate the representation  
 of  the geometrical information of 
the universe in terms of spectra, i.e. a set of eigenvalues of 
the Laplacian defined on the universe. Here, we concentrate only 
on one specific problem along this line: To 
 introduce a concept of distance  between universes 
in terms of the difference in the spectra.

 We can find out  such a measure of closeness from a general discussion:
 First, we introduce a suitable functional $P_{\cal G}[\cdot]$, where
  the geometrical information ${\cal G}$
    (represented by the spectra)  
  determines the 
   detailed shape of the 
  functional. Then, the overlapping functional 
  integral between $P_{\cal G}[\cdot]$ 
  and $P_{\cal G'}[\cdot]$ is  taken, 
  providing a measure of closeness between 
  $\cal G$ and $\cal G'$, $d({\cal G}, {\cal G'} )$. 
  
  The basic properties of this distance  (hereafter 
  referred to as `spectral distance', for brevity) 
   are then investigated. First, it can be related  
   to a reduced density 
   matrix element in quantum cosmology 
   between $\cal G$ and $\cal G'$. 
    Thus, calculating the spectral distance 
   $d({\cal G}, {\cal G'} )$ gives us an 
   insight for the quantum theoretical decoherence between two universes, 
    corresponding to $\cal G$ and $\cal G'$. Secondly, 
    the spectral distance becomes divergent except for when 
   $\cal G$ and $\cal G'$ have the same  dimension and volume. This is 
   very suggestive if the above-mentioned density-matrix interpretation is 
   taken into account. Thirdly, $d({\cal G}, {\cal G'} )$ does not 
   satisfy the triangular inequality, which illustrates clearly that 
   the spectral distance and the distance defined by the DeWitt metric
    on the superspace are not equivalent. 
    
   We then pose a question: 
   Whether two universes with different topologies 
   interfere with each other quantum mechanically? In particular, 
   we concentrate on the difference in the orientabilities. 
   To investigate this problem, 
   several concrete models in 2-dimension are set up, 
   and the spectral distances between them are investigated: 
   Distances between tori and  Klein's bottles, and those between 
   spheres and real projective spaces.
   Quite surprisingly, we find many cases of spaces with 
   different orientabilities in which 
   the spectral distance turns out to be very short. 
   It may suggest that, without any other special mechanism,  
    two such universes interfere with each other quite strongly,
     contrary to our intuition. 
   
   We discuss some curious 
   features of  the heat kernel for  tori and Klein's bottles in terms 
   of Epstein's theta and zeta functions.
   Differences and parallelisms between the spectral distance and 
   the DeWitt distance are also discussed.  
   }



\chapter{Introduction}

There are many situations  
in spacetime physics in which space/spacetime topology takes part.
First of all, it is a central problem of modern
 cosmology to determine whether 
 our universe is open or closed (i.e. non-compact or compact) [1] and 
 what kind of   topological structure  our universe possesses [2],[3],[4].
 The so-called worm-holes (topological handles attached to the universe)
 can cause many interesting phenomena, e.g. geons [5], 
 charge without charge [6], time-machines [7]. 
 The phenomenon of the topology change is one of the most intriguing problems 
 in quantum cosmology [8].
  
 Amongst these phenomena, 
let us now look at the scale-dependent topology 
(or physical topology) [9], [10] in more 
detail. Mathematically, topology can be looked upon as a global property of 
a manifold classified by  the concept of continuity or continuous 
deformations [11]. Thus by definition, it is a scale-independent concept,  
i.e. 
  objects with different scales, but continuously 
being deformed to each other are identified. 

However, once the concept of 
topology is applied to spacetime physics, 
the situation becomes different [10]:
Because of the limitation in the observational energy scale, thin 
objects smaller than $E^{-1}$ cannot be observed ($E$: the energy 
scale of observation). Suppose the universe is topologically 
complicated, with two kinds of small topological handles
\footnote\star{
The term `handle'  usually means an object diffeomorphic to 
$D^r \times D^{n-r}$ ($0 \le r \le n$), where $D^k$ stands for 
$k$-dimensional disk, and $n$ is the dimension of a manifold.
In this paper, however, any object which is topologically non-trivial 
 is generally called  `handle'.  }
 attached, 
 one kind  of scale $l$ and the other kind  of scale $L$, 
$l \ll  L$. If our observational energy is between $L^{-1} $ and 
$l^{-1}$, the smaller handles cannot be observed and the 
effective topology becomes simpler than the original one. 
If the energy scale is decreased further, such that it becomes
 less than $L^{-1}$, 
 bigger handles also cannot be observed, resulting in a much simpler 
effective topology. Such a picture of spacetime structure, that the  
real universe is topologically very complicated, but the effective 
topology becomes simpler when the observational energy scale becomes 
lower, originates from Wheeler's spacetime foam picture [12]. 
In short, the scale-dependent topology (physical topology) can be looked 
upon as topology with distinction between big and small handles.
 
 There is a clear lack of  a suitable language  to describe such 
 phenomena sketched in the previous paragraph. 
 To describe a sequential change in the effective topology of 
 a space as a function of observational energy,
 we first need to formulate closeness between two spaces 
 whose geometries are   different globally as well as locally. 
 There is no known 
 mathematical theory suitable for this purpose. The aim of this paper 
 is to formulate the concept of closeness between two spaces 
  from the viewpoint of spacetime physics.

  The main obstacle is the fact that we have to represent information 
  of topology as well as local  geometry in a quantitative, unified 
  manner. Let us then pay attention to the idea of using the 
  spectra  to characterize the geometrical content of 
   the universe: Basic vibrations (harmonics) of some  matter field on 
   the universe should  reflect the local and global 
   geometry of the universe.  
  For definiteness, let $({\cal M}, g)$ be a compact 
  Riemannian manifold. Here $g$ is a positive definite metric.
 Therefore, $({\cal M}, g)$ is regarded either the spatial section of 
the $(n+1)$-dimensional universe, or the Euclidean $n$-dimensional 
universe.
As the simplest elliptic operator, take the Laplace-Beltrami operator 
  $\Delta = g^{ab}D_a D_b 
  = {1 \over \sqrt{g} } \partial_a (\sqrt{g} g^{ab} \partial_b )$,
   where $D_a$ stands for the covariant derivative operator.  
   Then, we can set up  the eigenvalue problem on $({\cal M}, g)$;  
   $\Delta \psi + \lambda \psi =0$ with a suitable boundary condition.
 For simplicity, let us 
 assume $\partial {\cal M} = \emptyset $ here. 
Then, a set of eigenvalues is  obtained; 
$\{ \lambda_k \} $, $0=\lambda_0 < \lambda_1 \le \lambda_2 \cdots 
\uparrow \infty $. This set of  eigenvalues of the 
Laplacian clearly contains the information of 
 both local and global geometry of $({\cal M}, g)$. Hereafter, 
 these eigenvalues of the Laplacian shall be  called {\it spectra}, 
 for brevity.
  We may be able 
 to treat   these spectra  as a part of the fundamental  quantities  of 
 spacetime physics. 
 
The above idea of representing the geometrical content of the universe
 in terms of 
spectra is deeply related to a well-known mathematical problem, which is 
usually stated as `Can one hear the shape of a drum?' [13], i.e. 
a problem asking as  to what extent  
spectra  reflect the underlying geometry.
 In this spirit, the motto of our approach may be stated as 
 `Let us hear the shape of the universe.' Here, the appearance of 
 two terms `hear' and `shape' symbolizes the interaction between physics and 
 mathematics. In short, our idea is to convert the space(time) structure 
 (a mathematical object) into spectra or 
 `components of sound' (physical objects). 
This conversion may  be called  `spectral representation' of 
the spacetime structure. 

There are several advantages of such a representation. 
First, spectra represent the information of local and global geometry 
in an unified manner, which is suitable for  applications of quantum 
gravity/cosmology. Secondly, the spectral representation  is 
 a representation of the geometry 
in terms of  a countable set of 
real, positive numbers, which is easy to handle.
 Thirdly, spectra are the 
  diffeomorphism invariant quantities, 
 which is appropriate from the viewpoint  of general relativity.
 
 However, we should keep the following point in mind from the very 
 beginning: There exist the isospectral manifolds, i.e. 
  two Riemannian manifolds which are non-isometric to each 
 other,  but 
 have  identical spectra. Such a case has first  been  constructed by 
 Milnor on $T^{16}$ [14], and  several other cases have also been 
 presented later [15],[16],[17].
  Thus, in general, the information contained in geometry is 
 larger than the one represented by spectra. It is uncertain, however, 
 as to what extent these counter-examples are influential on  
 spacetime physics. In any case, it is clear that 
 the idea of the spectral representation is still worth while to pursue 
 extensively. 
 
 In this spirit, we shall concentrate on one specific, tractable  
  problem, which has been already implied above, i.e. 
   the introduction of a  measure of closeness 
 between two universes of different geometrical structures.

In section 2, we shall construct a general theory of 
the `spectral distance', i.e.  a measure of 
closeness between two manifolds in terms of spectra. 
Then, the physical interpretation of this spectral distance 
shall be investigated. Afterwards, 
the necessary condition for  the convergence
 of the spectral distance shall be studied. 
 We shall also define a scale-dependent 
 spectral distance, which provides a quantitative description of the 
 scale-dependent topology. Finally, we shall discuss  
 a peculiar property of the spectral distance,  i.e. 
 the failure of the triangular inequality. This result clearly 
 illustrates the non-equivalence between the spectral distance and the 
 distance defined by the DeWitt metric on the superspace [18].
 
In section 3, we shall pose a quantum cosmological question as to 
whether universes with different topologies (especially different 
orientabilities) interfere with each other 
quantum mechanically. We shall use the spectral distance as a 
suitable tool for analyzing this question. To investigate 
this problem, we shall 
 construct concrete 2-dimensional models of $T^2$ and  Klein's bottles, 
 and those of 
 $S^2$ and ${\bf R}P^2$.  We shall then calculate the spectral distances 
 between them. 
 We shall find many cases in which  the spectral distance between 
 two manifolds of different orientabilities turns out to be 
  surprisingly short.   This result may imply that two universes with 
  different orientabilities sometimes interfere quite strongly, contrary to 
  our intuition. The DeWitt distances shall be 
   also calculated for the 
  cases of $T^2$ and they are compared with the spectral distances. 
  We shall also construct and investigate the 
  heat kernel for   $T^2$ and  Klein's bottles in terms of 
  Epstein's theta and zeta functions [19],[20],[17].
  
In the final section, the spectral distance and the distance 
defined by the DeWitt metric  shall be 
 compared with each other in detail. 
Several other discussions shall  also be presented.




\chapter{The spectral distance and its properties}


\section{ The spectral distance}

In this subsection, we shall  search for 
the measure of closeness between two given manifolds.
 Let $({\cal M}, g)$ be a $n$-dimensional Riemannian manifold. For 
 clarity of discussion, we shall assume  that $({\cal M}, g)$ is compact
  and $g$ has the Riemannian signature, and not the pseudo-Riemannian
   signature. (Thus, $({\cal M}, g)$ can be regarded as a mathematical 
   model of the spatial section of the $(n+1)$-dimensional universe, 
   or a model of the Euclidean $n$-dimensional universe. 
   Hereafter, we may sometimes 
   refer to  $({\cal M}, g)$  just as `space' or `universe', 
   depending on the context.)
    The manifold $({\cal M}, g)$ includes two types of information:
   local geometry and global topology. We shall use the term 
   {\it geometry} in the broadest sense, including both. We shall 
   also use the symbol 
   $\cal G$ to represent the geometrical information 
   contained in $({\cal M}, g)$ in this broad sense.

Now, suppose two geometries $\cal G$ and $\cal G'$ are given, corresponding 
to manifolds $({\cal M}, g)$ and $({\cal M'}, g')$, respectively. 
Our goal is to find out a suitable formula for the `distance'
 $d({\cal G}, {\cal G'})$, representing  a closeness between 
 $\cal G$ and $\cal G'$ in terms of the eigenvalues 
(hereafter {\it spectra}, for brevity ) of some elliptic 
 differential operator $D$.
 From the viewpoint of 
 physical applications, the most interesting and simplest choice is 
 to take as $D$ the Laplace-Beltrami operator $\Delta$, 
 $\Delta = g^{ab}D_a D_b 
  = {1 \over \sqrt{g} } \partial_a (\sqrt{g} g^{ab} \partial_b )$,
   where $D_a$ stands for the covariant derivative operator.
\footnote\star{
Another interesting choice as $D$ may be the Dirac operator. 
In this paper, however, let us confine ourselves to the case of 
the Laplacian, since it is more basic and well-investigated in geometry.
}
 The basic idea for doing this 
 is as follows: 
\item{ (1)} 
Instead of comparing geometries directly 
(which is difficult),  introduce a suitable functional 
$P_{\cal G} [\cdot]$,
where the information of  a geometry $\cal G$ enters into this functional 
as a parameter and reflects on the  shape of the functional. 
\item{(2)} 
By  taking the overlapping functional integral between   
 $P_{\cal G} [\cdot]$ and $P_{\cal G'} [\cdot]$, we can indirectly 
 measure the closeness of two geometries ${\cal G}$ and ${\cal G'}$.

As a preliminary consideration, let us deal with  
functions instead of functionals: Suppose we need to introduce 
a concept of closeness in a set ${\cal S} = \{a,b,\cdots \}$. 
First we fix a function in which each element of ${\cal S}$ enters as 
a parameter, $\{ p_a(\cdot), p_b(\cdot), \cdots   \}$. (For instance, 
${\cal S}$ is a set of positive numbers and $p_a(\cdot)$ is a  Gaussian
 function with dispersion $a$.) 
Let us choose   functions $p_i(\cdot)$ $(i=a,b,\cdots)$ such that 
 $p_i(\cdot) \ge 0$ and $\int_{-\infty}^{\infty} p_i(x) dx = 1$.  
 In this case,  a suitable overlapping integral is 
$ \xi _{ij}=\int _{-\infty}^{\infty} (p_i(x) p_j(x))^{1\over 2} dx$, 
 which has the properties,

\item{({\rm i})} $  \xi _{ij}=\xi _{ji}$.  
\item{({\rm ii})} $0 \le \xi _{ij} \le 1$,\ \ \  and 
\item{}   $ \xi _{ij} =0 \iff p_i(\cdot) p_j(\cdot) \equiv 0$,\ \ \  
 $ \xi _{ij} =1 \iff  p_i(\cdot) \equiv p_j(\cdot)$.

Thus, we can define a measure of closeness between `$i$'  and  `$j$' 
as 
$$
d(i, j) = - \ln \xi_{ij} 
= -\ln \int _{-\infty}^{\infty} (p_i(x) p_j(x))^{1\over 2} dx\ \ \ .
$$
Then, $d(i, j)$ takes some value between $0$ (complete overlapping)
 and $\infty$ (complete non-overlapping). Here ends this preliminary 
 consideration.

There are several possibilities for the choice of $P_{\cal G} [\cdot]$
 and the form of the overlapping functional 
integral. Here, we are mainly interested in 
 the physical applications of the distance, so that we have to select  
 a `physically interesting' distance amongst  other possibilities.

First, we clearly need a  distance 
$d({\cal G}, {\cal G'})$ which is diffeomorphism invariant w.r.t.  
${\cal G}$ and ${\cal G'}$. As a  simple 
functional which is diffeomorphism invariant,  which reflects 
the global  as well as the local geometry,  and which includes 
the derivative operator (since it should be related to the spectra of 
a  differential operator), 
we take $\sigma [\cdot] $ as
\footnote\star{
We can also choose as 
$\sigma[f]:={1\over 2}\int_{\cal M} \{ (\partial f)^2 + \mu^2 f^2 \}$.
Here, $\mu$ is a smooth function.
 Then, the spectra are shifted, but there is no essential difference. 
 Therefore, let us assume  the simplest form, Eq.(1).} 
$$
\sigma[f]:={1\over 2}\int_{\cal M} (\partial f)^2
 ={1\over 2}\int_{\cal M} g^{ab}\partial_a f \partial_b f \sqrt{g}\ \ \ .
\eqno(1)
$$
Roughly speaking, in the above functional, 
the appearance of the metric $g$ reflects the local 
geometry, 
 while the integral over $\cal M$ reflects the global geometry.

However, $\sigma[\cdot]$ is not enough for our purpose since it 
takes an arbitrarily large positive value so that it cannot be normalized. 
Thus, let us fix  a suitable function $F(\cdot)$ such that 
$P[\cdot]:= F(\sigma[\cdot])$ is well-behaved. In view of the preliminary 
consideration above, $P[\cdot]$ should satisfy
\item{ ({\rm I})} $P[\cdot] \ge 0, \ \   \int [df]\ P[f] = 1\ \ \ .$

Furthermore,  
we shall 
need a physically sensible `distance' applicable to the analysis of the 
scale-dependent topology, which requires  that smaller handles should  be
 ignored under 
  certain conditions,  compared with larger handles. Thus,
\item{({\rm II})} The higher (lower) weight for the longer (shorter) scale 
           behavior of $\cal G$.

  There are still various possibilities that remain for the choice of 
  $F(\cdot)$. Here, for several reasons, 
  we shall choose one  possibility amongst many others, 
  $F(x)= \exp -x$: First, it is a simple function which is 
  easy to handle. Secondly, it makes us possible to relate 
  the distance with a physical concept, i.e. the quantum decoherence 
  of the universes, which shall be discussed in detail 
 in the next subsection. 
  Thirdly, it satisfies the requirements (I) and (II):
   Consider the eigenvalue problem
   $\Delta \phi_k + \lambda_k \phi_k =0 $ (with a suitable 
    boundary condition;  mainly 
   the case  of $\partial {\cal M} = \emptyset$ shall be described here.)
    Thus we get the spectra
   $0= \lambda _0 < \lambda_1 \le \lambda_2 \cdots \uparrow \infty \equiv 
   \{ \lambda_k \} $. If $f$ is expanded in terms of the eigenfunctions 
   $\{  \phi _k \}$, 
   $f(\cdot)=\Sigma_{k=0}^{\infty} \alpha _k \phi _k (\cdot)$, then, 
 $\sigma_{\cal G}[f]= {1\over 2} \Sigma_{k=0}^{\infty} 
 \lambda _k \phi _k ^2$. By choosing $F(x) = \exp -x$,  thus, 
  $P_{\cal G}[f] = 
  \exp -{1\over 2} \Sigma_{k=0}^{\infty} \lambda _k \phi _k ^2$. 
  This satisfies (I) obviously by choosing a suitable measure for 
  the functional integral. Noting that the smaller (larger) 
$\lambda_k$ in $\{ \lambda_k \}$ coarsely reflects the larger (smaller) 
scale behavior of $\cal G$,  the above choice also satisfies (II). 
(Compare two values of $P_{\cal G }[f]$ for,
 say, $f \sim \phi _3$ and $f \sim \phi _{100}$.)

Now, we shall 
formally generalize the procedure in the preliminary consideration. 
First, from the normalization requirement of $P_{\cal G}[\cdot]$ 
in (I), we find 
$$
[df]=\prod_{k=1}^\infty \sqrt{  {\lambda_k \over 2 \pi}   }d \alpha_k
\ \ \ ,
$$ 
where the zero-mode $\lambda_0$ is understood to be  removed if it appears 
(e.g.  the case of $\partial {\cal M} = \emptyset$). Formally repeating 
the same procedure as in the preliminary consideration, then, 
$$
\eqalignno{
\xi ({\cal G}, {\cal G'})&=\int [df] (P_{\cal G}[f] P_{\cal G'}[f])^{1\over 2} 
                                \cr
 &=\int \prod_{k=1}^{\infty} { d\alpha _k \over (2\pi)^{1\over 2} }
             (\lambda_k {\lambda'}_k)^{1\over 4} 
              \exp -{1 \over 4} \Sigma_{m=1}^{\infty}(\lambda_m + {\lambda'}_m)
                                                \alpha_m ^2 \cr
       &=  \prod_{k=1}^{\infty} 
       \left\{ 
       {1\over 2} 
        \left(\sqrt{\lambda_k \over {\lambda'}_k} + 
        \sqrt{{\lambda'}_k \over \lambda_k} \right)
        \right\}^{ -{1\over 2} }\ \ \ . 
  & (2) \cr
   }      
$$
Thus, 
$$
\eqalignno{
d({\cal G}, {\cal G'})&= -\ln \xi ({\cal G}, {\cal G'}) \cr
                     &={1\over 2}\Sigma_{k=1}^{\infty}
                     \ln {1\over 2} \left(\sqrt{\lambda_k \over \lambda_k'} + 
        \sqrt{\lambda_k' \over \lambda_k}  \right)\ \ \ . 
    & (3) \cr    
}                                    
$$
For brevity, let us refer to  this quantity as the 
`spectral distance' between $\cal G$ and $ \cal G'$.

Finally, let us note one subtlety included in the above procedure
 for deriving Eq. (2): In taking the overlapping functional integral,
 $\int [df] (P_{\cal G}[f] P_{\cal G'}[f])^{1\over 2}$, 
 a prescription for the identification of a function $f$ defined on 
 $\cal G$ with  {\it another}  function $f$ defined on $\cal G'$ 
should be fixed.
 Here, we have chosen the following prescription: 
 Let $\cal G$ and $\cal G'$ be compact Riemannian manifolds, and 
 $D$ be an elliptic differential operator. Let 
 $ 0= \lambda _0 < \lambda_1 \le \lambda_2 
 \cdots \uparrow \infty \equiv  \{ \lambda_k \}$ and 
 $ 0= \lambda' _0 < \lambda'_1 \le \lambda'_2 
 \cdots \uparrow \infty \equiv  \{ \lambda'_k \}$ 
 are spectra of $D$ on  $\cal G$ and $\cal G'$, 
 respectively, numbered in the  increasing order. Then, 
 we have sets of eigenfunctions of $D$, $\{  \phi _k (\cdot) \}$ and 
 $\{  \phi' _k (\cdot) \}$, corresponding to $\{ \lambda_k \}$ and 
 $\{ \lambda'_k \}$, respectively. A function $f(\cdot)$ 
 on $\cal G$ can be expanded in terms of $\{  \phi _k (\cdot) \}$, 
 $f(\cdot)= \sum_k \alpha_k \phi_k (\cdot) $. Then, our  prescription is that 
 $f(\cdot)$ on $\cal G$ is identified w.r.t. $D$ with a function
  $\sum_k \alpha_k \phi'_k (\cdot)$ on $\cal G'$.
  
 In  cases when $\cal G$ and $\cal G'$ are within some 
  one-parameter family of geometries, this prescription may be 
  justified rigorously   through the  adiabatic theorem [21]. Therefore, 
  this way of identification  seems to be the most natural one. 
  This subtlety of identifying functions on different spaces also 
  emerges in quantum field theory on a curved spacetime. The same  
  prescription is implicitly adopted in this case 
  (See the next subsection  for more details). 
  Even if  there is  
  a better prescription, the basic idea and procedures  remain same, and 
  only the final result Eq.(2) would be subject to some modification.
  Here, let us adopt the above-mentioned prescription.

\section{ The physical interpretation of the spectral distance}


 Physically, it is natural 
 to expect that the geometrical information reflects on the behavior 
 of a field distributed on $({\cal M},g)$.
 In fact, the functional $\sigma[\cdot]$ (Eq.(1)) is in the form 
 of the action of a  scalar field on $({\cal M},g)$. (Note, 
 however, that $({\cal M},g)$ is Riemannian.)
 Thus, one may  suppose  
 that the spectral distance $d({\cal G}, {\cal G'})$ yields a 
 physical interpretation. Indeed,  the quantity 
 $\xi ({\cal G} ,{\cal G'}) = \exp -d({\cal G}, {\cal G'})$ (Eqs.(2)
and (3))  turns out to be  related  to 
  a  reduced density matrix element for the universe, which appears in 
  the discussions of the emergence of the classical world from the 
  quantum universe [22],[23],[24]. Let us now see this point in order 
  to make clear the assumptions behind this coincidence, i.e. the 
  relation of $d({\cal G}, {\cal G'})$ with a reduced density 
  matrix element.
  
  Let us consider the system of gravity and a massless scalar field.
  \footnote\star{
  There is no essential difference between the massless case and the 
  massive case. }
   It is interesting to consider other kinds of fields 
  also, the Dirac field for instance. However, 
  the structure of the vacuum state for a fermion field 
  looks quite different from the one for  a boson field, at least 
  mathematically, and requires  separate investigations. To avoid 
  extra complications, therefore, let us consider only a scalar field 
  here ({\it Assumption} $a$).
The total action is given by 
  (the signature 
 is chosen as $(-,+, \cdots, +)$), 
 $$
  \eqalign{
 S[{\cal G}, \phi] &={1/\alpha.}\int R \sqrt{-g}
  + \int (-{1 \over 2} \partial_a \phi \partial^a \phi 
  -{1\over 2}m^2 \phi^2) \sqrt{-g} \cr
  & =:\ S_{grav}[{\cal G}] + S_{matter}[{\cal G}, \phi]\ \ , \cr
}
  $$
 where $\alpha$ is a suitable gravitational constant. Here, 
 the spatial geometry $\cal G$  and the scalar field 
 $\phi$ induced on $\cal G$ are the configuration variables.  
This system obeys the quantum theory, described by the Wheeler-DeWitt
 equation [18], $H \Psi [{\cal G}, \phi] = 0$, where $H$ is the
Hamiltonian constraint obtained from $S[{\cal G}, \phi]$. In the semiclassical 
 region, it may be  a good approximation to do the quantum theory 
 separately for $S_{grav}({\cal G})$ and $S_{matter}[{\cal G}, \phi]$:
On the one hand, we regard that the dynamics of ${\cal G}$ is 
approximately described solely by $S_{grav}[{\cal G}]$; On the other hand, 
$\phi$ is described by $S_{matter}[{\cal G}, \phi]$, with ${\cal G}$
 treated just as parameters ({\it Assumption} $b$).
 \footnote\star{
 Usually, this treatment is regarded as valid on grounds of the 
 `smallness' of $\alpha$. More rigorously, the typical amplitude of 
 quantum fluctuations of spacetime 
 should be taken into considerations for the 
 justification of this treatment [4].
 }
  Thus we set the Ansatz for 
 $\Psi ({\cal G}, \phi)$ as, 
 $\Psi ({\cal G}, \phi)= \varphi({\cal G}) \cdot \eta({\cal G}, \phi)$.

Now, let us consider $\eta[{\cal G}, \phi]$, described by 
$S_{matter}[{\cal G}, \phi]$. Choosing $N=1$, $N_i =0$, $S_{matter}$
becomes, 
$
S_{matter}[{\cal G}, \phi] =
{1\over 2}\int \left\{ {\dot \phi}^2 - \phi (-\Delta + m^2)\phi \right\}
\sqrt{h}
$ ($h:= det\ h_{ab}$, $h_{ab}$ is the induced spatial  metric). 
It is natural  to expand 
$\phi (t, {\vec x})$ in terms of the normalized eigenfunctions,  
$\left\{ \phi _k ({\vec x}) \right\}$,  
of the elliptic operator $D:= -\Delta + m^2$, 
 satisfying   eigenvalue equations 
$D \phi_k (\cdot)=\lambda_k ({\cal G})  \phi_k (\cdot)$, with a 
suitable boundary condition compatible with the action principle for 
$S_{matter}[{\cal G}, \phi]$.  Thus, 
$\phi (t, \cdot) = \sum _k \alpha _k (t) \phi _k (\cdot)$.  Then, 
$S_{matter}[{\cal G}, \phi]= \sum_k {1\over 2} 
\int  (\dot{ \alpha_k} ^2 - \lambda_k ({\cal G}) \alpha_k ^2) dt$, 
which is equivalent to a collection of harmonic oscillators 
(note that $\cal G$ is now treated as a 
fixed parameter). 
\footnote\S{
It is notable that $\phi$ appears  in the theory only through 
 a set of  expansion coefficients 
$\left\{ \alpha_k \right\}$ w.r.t 
$\left\{ \phi _k (\cdot) \right\}$, a set of eigenfunctions 
for $D$. In this sense, $\left\{ \alpha_k \right\}$ can be regarded 
more fundamental than $\phi$ itself.}
Afterwards, it is in principle straightforward to quantize this system.
 Technically, the simplest way to quantize it is to choose 
 the adiabatic vacuum as quantum state ({\it Assumption} $c$): We assume that
  the typical time scale of the change    in the field $\phi$ is  much 
     shorter  than the one of the change in geometry $\cal G$. Then we  
     define  the vacuum state at each instant of time;
     $$
     \eta ({\cal G}, \phi)=\prod_k \left(
     {\lambda_k({\cal G}) \over 2 \pi}\right)^{1/4} 
     \exp -{1\over 4} \lambda_k({\cal G}) \alpha_k^2\ \ \ .
     $$
The density matrix of the universe may be defined as  
$$\rho({\cal G}, \phi ; {\cal G'}, \phi ')
 = \Psi ({\cal G}, \phi) \Psi ^* ({\cal G'}, \phi')\ \ \ .
 $$ To give the meaning to  this density matrix, the wavefunction 
 $\Psi ({\cal G}, \phi) $ should yield the probabilistic interpretation.
 It means that $\Psi ({\cal G}, \phi) $ should be understood as 
 normalized w.r.t. a suitable inner product, or some alternative 
 interpretation for $\Psi ({\cal G}, \phi) $ should be provided 
 ({\it Assumption} $d$). 
 To discuss 
 the quantum decoherence of the universe, one may treat $\cal G$ as 
 the system variable and $\phi$ as the environment, and take the 
 partial trace w.r.t. $\phi$ (remember the Ansatz for 
 $\Psi ({\cal G}, \phi)$):
 $$
\eqalign{
 \rho_{reduced} ({\cal G},{\cal G'}) & =
 \int [d\phi] \rho({\cal G}, \phi ; {\cal G'}, \phi )
=\varphi ({\cal G})\varphi^* ({\cal G'}) 
\int [d\phi] \eta({\cal G}, \phi) \eta^* ({\cal G'}, \phi) \cr
& =:\ \varphi ({\cal G})\varphi^* ({\cal G'}) \xi ({\cal G}, {\cal G '})
    \ \ \ ,\cr
}
 $$
 where $\xi ({\cal G}, {\cal G '})$ exactly agrees with  the one 
 given by Eq.(2), the latter being  expressed as 
 $\xi ({\cal G}, {\cal G '})= \exp -d ({\cal G}, {\cal G '})$. 
 Therefore, in our terms, the longer the spectral distance 
 $d ({\cal G}, {\cal G '})$ is, the smaller the corresponding 
 off-diagonal element $\rho_{reduced} ({\cal G},{\cal G'})$ is, 
 implying the stronger decoherence between the universes 
 ${\cal G}$ and ${\cal G'}$.

 In taking the partial trace w.r.t. $\phi$, the operation 
 $\int [d\phi] \eta({\cal G}, \phi) \eta^* [{\cal G'}, \phi] $
 should be of meaning. This procedure  contains two subtle points: 
 First, it corresponds to comparing states for 
 two different harmonic oscillators with different frequencies 
 ($\lambda_k ({\cal G})$ and $\lambda_k ({\cal G'})$). 
 Mathematically, it is just a change of basis in a functional space.
  However, physically it presupposes the identification of two 
  Hilbert spaces, one characterized by $\lambda_k ({\cal G})$, and 
 the other by $\lambda_k ({\cal G'})$. This is   not 
 within the  framework of ordinary  quantum theory.
 \footnote\star{
 One situation in usual  quantum theory, 
 which is of some similarity with the present problem, appears in 
 the discussions of the adiabatic perturbations [21]. 
 In this case, the Hilbert 
 space for the system characterized by parameters, say $\omega $, can be 
 asymptotically related with  
 the other Hilbert space characterized by $\omega'$,  
 when  $\omega \rightarrow  \omega'$. However, the problem of  our 
 concern requires even the comparison of two drastically different 
 Hilbert spaces.
 }
  Secondly, 
  a reasonable prescription to identify 
  a function $\phi$ on $\cal G$ with 
  the same on $\cal G'$ should be fixed. The rule adopted here is  
  to identify $\sum _k \alpha _k  \phi _k (\cdot)$ with 
  $\sum _k \alpha _k  {\phi'} _k (\cdot)$. Here, 
 $\left\{ \phi _k (\cdot) \right\}$ and 
 $\left\{ {\phi'} _k (\cdot) \right\}$ are   eigenfunctions 
of $D$ on $\cal G$ and $\cal G'$, respectively.
  Thus, we face with the same  subtlety of  identifying functions 
  defined on $\cal G$ and $\cal G'$ as
   discussed in \S\S 2-1.
 These subtleties  always emerge,  explicitly or implicitly, whenever 
 we discuss the  quantum field theory in curved spacetime. 
 All we can do at present is to expect that once the inner product 
 in a space of wavefunctions of the universe, $\Psi ({\cal G}, \phi)$, 
 shall be  correctly fixed, and a complete interpretation of  
 $\Psi ({\cal G}, \phi)$  shall be given, this problem would be 
 automatically solved ({\it Assumption} $e$). (Therefore, 
 {\it Assumption}s $d$ and  $e$ are deeply linked with each other.)
 
Ref.[22] may be one of the earliest works which pays  attention to the 
reduced density matrix for the explanation of the classical behavior
of the quantum universe. Ref.[23] pursues this idea explicitly 
for the case of the perturbed Friedmann universe with a massive scalar
field, based on the model of Ref.[25]. In this case, 
unperturbed quantities , i.e. the scale factor $a(t)$ and the 
homogeneous component  of the scalar field, $\phi (t)$, correspond to 
our $\cal G$. For the tensor modes of perturbations, for
instance, one can essentially set $\lambda_k = {k \over 2}a^2$. Then the
suppression factor 
$\xi (a, a')= \lim _{N \rightarrow \infty} \exp -{N\over {4aa'}}(a-a')^2$ 
appears, which agrees with Eq.(2), with 
 $\lambda_k = {k \over 2}a^2$.
   \footnote\star{
   More precisely, because $\phi (t)$ is also included 
   in the category of our $\cal G$, the combination \nextline
   ${k \over 2}a^2 -i2a^3 m \phi$ is to be identified with $\lambda_k$, 
   yielding the factor \nextline
   $\xi (a, \phi ; a', \phi' )
   = \lim _{N \rightarrow \infty} \exp -{N\over {4 aa'} }(a-a')^2
    \exp - {\pi^2 \over 4} m^2 a a'(\phi-\phi')^2$ [23].
   }
 Ref.[24] recapitulates the arguments of Ref.[22] and [23] 
in more general terms. It even implies the idea of the distance,
though no further investigations  (e.g. its convergence condition,
comparison with the axioms of distance) are pursued. Needless to say, 
these works, [22-24], are all subject to the assumptions and subtleties 
pointed out above.

Let us summarize the physical interpretation of the spectral distance
 $d({\cal G}, {\cal G'})$ introduced in \S\S 2-1: 
 The  quantity $\xi ({\cal G}, {\cal G'}) 
 = \exp -d({\cal G}, {\cal G'})$ boils down to 
  a  reduced density matrix element (for a system of the universe 
  ($\cal G$) and  a scalar  field ($\phi$), with $\phi$ being 
  traced out) under 
 the following conditions:
 \item{(a)} A suitable 
 inner product between  the wavefunctions of the universe, 
   $\Psi({\cal G}, \phi)$ and 
 $\Psi({\cal G'}, \phi')$ can be defined, so that the density matrix 
  admits quantum theoretical interpretation.
  \item{(b)} The spacetime can be treated semiclassically, 
  and the scalar field  $\phi$ on $\cal G$ and  $\cal G'$ is 
  in the ground state and 
  can be treated  adiabatically (the time scale of the change 
    in the field is regarded as much 
     shorter  than the one of the change in geometry). 

This coincidence of $\xi ({\cal G}, {\cal G'})$ boiling down to the 
reduced density-matrix element 
suggests several important things: 

First, the specific choice of $F(x)= \exp -x$ in \S\S 2-1
 is clearly distinguished 
amongst  several possibilities, from the view of physical applications.

Secondly, once we  obtain  a satisfactory quantum  theory 
of gravity in the future to 
compute $\rho_{reduced } ({\cal G}, {\cal G'})$ exactly, it would 
be relevant to define `closeness' between $\cal G$ and $\cal G'$ 
as $-\ln \left\{ \rho_{reduced } ({\cal G}, {\cal G'})
 / \varphi ({\cal G})\varphi^* ({\cal G'}) \right\}$. 
 In other words, 
the stronger two geometries  interfere 
with each other quantum mechanically,  the `closer' they can be 
regarded (a suggestion from physics for mathematics).

Thirdly, one may be able to estimate 
$\rho_{reduced } ({\cal G}, {\cal G'})
/ \varphi ({\cal G})\varphi^* ({\cal G'})$ by calculating 
\nextline
$\exp - d({\cal G}, {\cal G'})$ according to (3) 
(a suggestion from mathematics for physics).
Needless to say, we have no satisfactory theory of quantum gravity 
at present, so that $\rho_{reduced } ({\cal G}, {\cal G'})$ cannot be 
calculated exactly. However, the reduction of  
 $ d({\cal G}, {\cal G'}) $ (which has been derived from 
general arguments and which is  applicable to any geometries in principle)
 into $- \ln \left\{ \rho_{reduced } ({\cal G}, {\cal G'})
 / \varphi ({\cal G})\varphi^* ({\cal G'})\right\}$ 
 (estimated approximately) 
 when ${\cal G}$ and ${\cal G'}$ are restricted to be of 
  the same dimensionality and topology, is very suggestive and should be 
  meaningful. 
  Thus, we can infer the following equality, 
$$  
d({\cal G}, {\cal G'}) 
= - \ln \left\{ \rho_{reduced } ({\cal G}, {\cal G'})
  / \varphi ({\cal G})\varphi^* ({\cal G'})\right\} \ \ .
$$

In connection with the condition $(a)$, we should note the following: 
The inner product between two wavefunctions defined on two different 
superspaces is much less known than the inner product between 
wavefunctions defined on the same superspace. 
(For instance, universes with different 
dimensions, or with different global topologies are subject to different
 superspaces.) In other words, there  is no established way of 
 comparing universes with different global structures. This is the 
 main obstacle which prevents us from undertaking extensive studies 
 on the phenomenon  of the topology change. At the same time, 
 this is one of the motivations of the introduction of the spectral 
 distance ($\S 1 $).  At present, we are usually forced to 
 restrict ourselves   
 to the comparably moderate  cases  in which universes 
 lie in the same superspace 
 (in most cases, minisuperspace or its generalization), assuming 
 somehow  the inner product.

Regarding the relation between the spectral distance and 
the reduced density matrix, 
therefore, one  had better  keep the following caveat in mind:
 Only for the  cases when ${\cal G}$ and  ${\cal G'}$ are subject to 
  the same superspace (typically, of the same dimension and topology),
    the above equality holds safely 
under  several assumptions. For other cases, it should be regarded 
as an extrapolation.

Since this caveat should   be always remembered whenever we shall connect 
the spectral distance with physical interpretations, 
we  refer to  it as  {\it Caveat A}, for brevity.  
It is quite probable that (some modified form of)  the above  equality 
turns out to be  generally true,  once a satisfactory theory of 
quantum gravity is  provided.

\section{The convergence condition of the spectral distance}

Let us now investigate the necessary condition for the convergence of 
the spectral distance,  since its definition in Eq.(3) includes 
an  infinite summation. 
Clearly, the {\it necessary}  condition for  convergence is 
$$
a_k({\cal G}, {\cal G'})
:={1\over 2} \left( \sqrt{\lambda_k \over {\lambda '}_k} + 
\sqrt{ {\lambda '}_k \over \lambda _k} \right) \longrightarrow 1 
\ \ {\rm as}\ \  
k \rightarrow \infty \ \ \ .
$$
 Note that $a_k \ge 1, \ \ = \iff \lambda _k = {\lambda '}_k$. 
 Thus the necessary condition 
for the convergence of $d({\cal G}, {\cal G'})$ is 
${ {\lambda '}_k \over \lambda _k} \rightarrow 1$ as $k \rightarrow \infty$.
 Now, we have  Weyl's asymptotic formula [26],[27], 
 which can be represented in several ways,
 $$
 \eqalignno{
 \Sigma_{k=0}^{\infty} \exp -\lambda_k t 
 & ={1 \over (4\pi t)^{n \over 2}} V + O(1)  
 \ \  {\rm  as}\ \      t \downarrow 0, & (4-a) \cr
 N(\Lambda) &\sim {  { {\bf \omega}_n V}\over   (2\pi)^n }\Lambda ^{n\over 2}
 \ \  {\rm  as}\ \  \Lambda \rightarrow \infty , & (4-b) \cr
 \lambda_k &\sim \left(  { {(2\pi)^n} \over {{\bf \omega}_n V} } k  \right)^
 {2\over n} \ \  {\rm  as}\ \  k  \rightarrow \infty,  & (4-c) \cr
 }
 $$
where $n=dim {\cal M}$=dimension of ${\cal M}$, 
$V=vol {\cal M}$=$n$-volume of $\cal M$, 
$N(\Lambda):= \# \{ \lambda_k| \lambda_k \le \Lambda  \} $ and 
${\bf \omega}_n$ := $n$-volume of unit $n$-disk in ${\bf R}^n$ 
(e.g. ${\bf \omega}_2 = \pi $).
 Significantly, the `$O(1)$-term' in $(4-a)$ reduces to 
 $\chi ({\cal M}) /  6 + O(t)$ 
 when $n=2$, $\cal M$ is  compact, and $\partial {\cal M} = \emptyset $, 
 where  $\chi ({\cal M})$ is the Euler number of $\cal M $ [26].
  We shall come across the application of this result in $\S 3$.  
Weyl's asymptotic formula $(4-c)$ 
states that the asymptotic behavior of $\lambda _k$ ($k \rightarrow \infty$)
 depends  on $dim {\cal M}$,  $vol {\cal M}$, and topology of $\cal M$,  but does not depend on 
  a detailed local geometry of $({\cal M}, g)$. From $(4-c)$,  
  $$
  { {\lambda'} _k \over \lambda _k} \sim 
  { { ({\bf \omega}_n V)^{2\over n}}  \over  
  { ({\bf \omega}_{n'}  V')^{2\over {n'}  } } }
                          k^{  {2\over {n'}}
                           -  {2\over n}  }
\ \  {\rm  as}\ \  k  \rightarrow \infty \ \ \ .
$$
Thus the necessary condition for  convergence of $d({\cal G}, {\cal G'})$
 is 
 $$
 dim\ {\cal M} = dim\ {\cal M'} \ \ \ {\rm and }\ \ \ 
 vol\ {\cal M} =vol\ {\cal M'}\ \ \ . \eqno{(5)}
 $$
This result is quite suggestive. According to the density-matrix
 interpretation
 with {\it Caveat A}, this suggests that two universes with 
 different dimension 
 or  volume decohere very strongly.
 
 Finally, we should note that Eq.(5) is just the necessary condition for 
 convergence,  but not a  sufficient one.

 \section{The scale-dependent spectral distance}

From the formula Eq.(3), we are naturally led to the scale-dependent 
spectral distance, by introducing a cut-off in the summation.
We shall compare the subsets of $\{ \lambda_k \}$ and  $\{ {\lambda'}_k \}$
 constructed from elements less than $\Lambda$. It corresponds to 
a coarse comparison of  two geometries $\cal G$ and $\cal G'$, neglecting 
the difference in the smaller scale behaviors  than 
$\Lambda ^{-{1\over 2}}$.

More specifically, let us define
$$
N_\Lambda :=Min \left( 
\# \{ \lambda_k \in \{ \lambda_k \}| 0 \le \lambda_k \le \Lambda  \}, \ \ 
\# \{ {\lambda'}_k \in \{ {\lambda'}_k \}| 0 \le {\lambda'}_k \le \Lambda  \}
 \right)\ \ \ .
$$
 In terms of $N_\Lambda$, we shall define $\Lambda^{\Lambda}_k$ as,
$$
\lambda^{\Lambda}_k 
:= \cases{ \lambda_k\ \ &{\rm  for\ \  $k \le N_\Lambda$ }\cr
   \lambda_{N_\Lambda} \ \ &{\rm  for\ \  $k > N_\Lambda$ } \ \ , \cr }
$$
then, $ \{ \lambda^{\Lambda}_k \}= \{ \lambda_0,
 \lambda_1, \cdots, \lambda_{N_\Lambda}, \lambda_{N_\Lambda}, \lambda_{N_\Lambda}, \cdots \}  $. In the same way, 
 we shall define 
$ \{ \lambda'^{\Lambda}_k \}= \{ \lambda'_0,
 \lambda'_1, \cdots, \lambda'_{N_\Lambda}, \lambda'_{N_\Lambda}, \lambda'_{N_\Lambda}, \cdots \}  $.
 Then, we shall define the scale-dependent spectral distance 
$d_{\Lambda}({\cal G}, {\cal G'})$ as
$$
d_{\Lambda}({\cal G}, {\cal G'})
= {1\over 2} \Sigma_{k=1}^{N_\Lambda} 
       \ln {1\over 2} 
\left( {\sqrt{\lambda^{\Lambda}_k \over {  {\lambda' }^{\Lambda}_k} } }+ 
\sqrt{ 
 { {\lambda' }^{\Lambda} _k \over {\lambda^{\Lambda} _k} }
 } 
 \right) \ \ \ .
\eqno{(6)}
$$ 

There are other possibilities in  the way of introducing a cut-off.
For instance, instead of taking the minimum of the two numbers 
in the definition of $N_\Lambda$, 
taking the  average of the two numbers is  one possibility. Replacing  
minimum  by maximum is another possibility. Weyl's asymptotic 
formula Eq.$(4-c)$ guarantees that the difference caused by different 
choices becomes negligible when $\Lambda$ is sufficiently large.

It is interesting to investigate the behavior 
$d_{\Lambda}({\cal G}, {\cal G'})$ as a function of $\Lambda$ for 
given $\cal G$ and  $\cal G'$. For instance, suppose a geometry 
$\cal G$  with a very complicated topological structure, 
${\cal G} = {\cal G'} \# h_1 \# h_2 \cdots \# h_m$, where $h_i$'s are 
some topological handles of a typical scale $l$. When the cut-off 
parameter $\Lambda$ is increased smoothly, 
the spectral distance $d_{\Lambda}({\cal G}, {\cal G'})$ is expected to 
increase  abruptly near $\Lambda \sim l^{-2}$, indicating  that
$\cal G$ and $\cal G'$ are almost similar in the scale larger than $l$, 
but they are very different in the scale smaller than $l$. This 
provides a new quantitative representation of the scale-dependent topology
 or topological approximation [10]. As already mentioned in $\S 1$, 
 the concept of scale-dependent topology or topological approximation 
 has many interesting applications.
  However, its rigorous quantitative 
 formulation is quite difficult: It requires the concept of 
 `closeness' between two topologically different Riemannian manifolds [10], 
 for which we have no mathematical theory as yet. To make this concept of 
 `closeness' or `distance' physically sensible, it should have some 
 connection with physical quantities. In this respect, the quantity 
 $d_{\Lambda}({\cal G}, {\cal G'})$ 
 is a good candidate for the  measure of closeness
  between $\cal G$ and $ \cal G'$, since it is defined in terms of spectra
  (where a matter field plays the role of 
  a probe for the geometrical structure of the universe)  
  and since it can be related to 
  the reduced density-matrix element for the universe (with {\it Caveat A})
   as discussed in \S\S 2-2. In Ref.[10], the quantitative description 
   of scale-dependent topology has been investigated 
   in terms of the scattering cross-sections, 
    treating topological handles as a scatterer. One restriction of this 
    framework is that it requires an asymptotic region 
    with  trivial topology to set up a scattering problem. On the 
    other hand, the description in terms of the scale-dependent 
    spectral distance does not assume such an asymptotic region.

\section{The failure of the triangular inequality}

  The ordinary requirements for a function $d: S \times S \rightarrow {\bf R}$ 
  ($S$: a set) to be regarded as a distance are
  \item{(I)} Positivity: $(a)$ $d(p,q) \ge 0$ \qquad 
           $(b)$ $d(p,q) = 0 \iff p=q$.
  \item{(II)} Symmetry: $d(p,q) =d(q,p)$.
 \item{(III)} The triangular inequality: 
        $d(p,q)+ d(q,r) \ge d(p,r)$.

The spectral distance $d({\cal G}, {\cal G'} )$ clearly satisfies 
$(I)-(a)$ and $(II)$. As for $(I)-(b)$, we know counter-examples
 which do not satisfy `$\Rightarrow$': There are examples of 
 a pair of non-isometric geometries ${\cal G}$ and ${\cal G'}$ whose 
 spectra are identical $\{ \lambda_k \} \equiv \{ {\lambda'}_k \}$.
 Two non-isometric geometries on $T^{16}$ with identical spectra
  have been given by Milnor [14]. Later, other examples have also been 
  constructed [15],[16] and they are called isospectral manifolds [17]. 
  (Note that the necessary conditions for ${\cal G}$ and ${\cal G'}$ 
  to be isospectral,  are  
 $dim\ {\cal M} = dim\  {\cal M'}$ and  
 $vol\  {\cal M} =vol\  {\cal M'}$ by Weyl's asymptotic formula 
 Eq.$(4-c)$.)
   Such a pair of isospectral 
   manifolds cannot be separated in terms of the spectral distance
    w.r.t. the Laplacian $\Delta$.
    \footnote\star{
    However, some other elliptic operators can yield the spectral 
    distance which distinguishes the isospectral manifolds w.r.t. 
    $\Delta$.} 
   These isospectral manifolds do not seem to be generic, so that 
     they may  be identified in the space of all geometries.
     On the identified space, $(I)-(b)$ can be regarded to hold good.
      This 
     observation implies that the spectral distance is a `coarser' 
     distance compared with the distance defined by the DeWitt metric [18] 
     (for manifolds with the same dimension, volume and topology). 
     Employing  the density-matrix interpretation of the spectral distance,
      it might imply that a pair of isospectral spaces interfere with 
      each other strongly. However,
      it is uncertain as to what 
      extent these exceptional cases have an influence 
       on the applications to cosmology.

  Finally, we investigate the validity of 
  $(III)$ the triangular inequality, 
  $d({\cal G}, {\cal G'}) + d({\cal G'}, {\cal G''}) 
    \ge d({\cal G}, {\cal G''})$.  
   We mention  in advance that in $\S 3$, where 2-dimensional models shall be 
   investigated, many cases shall  be found 
    in which  the triangular inequality does not hold. Thus, the spectral 
    distance does not satisfy the triangular inequality in general.
    The spectral 
  distance can be written as 
  $d({\cal G}, {\cal G'}) 
  = {1\over 2}\Sigma_{k=1}^{\infty} d_k({\cal G}, {\cal G'})$ 
  (see Eq.(3)).
   It is interesting to  investigate the condition for 
   the term-wise violation of the 
 triangular inequality:   
   First,  
   $$
  d_k({\cal G}, {\cal G'}) + d_k({\cal G'}, {\cal G''}) 
  = \ln {1\over 2} \{ {1\over 2}(1 + {1\over \alpha })\sqrt{\beta} 
         +  {1\over 2}(1 +  \alpha ) {1 \over  \sqrt{\beta} }   \},
   $$
   where $\alpha := \sqrt{\lambda_k \over {\lambda '}_k}$ and 
   $\beta := \sqrt{\lambda_k \over {\lambda ''}_k}$.  Next, 
   $$
  d_k({\cal G}, {\cal G''}) 
  = \ln  {1\over 2} (\sqrt{\beta} + {1 \over  \sqrt{\beta} })\ \ \ .
  $$\
  Now, $ {1\over 2}(1 +  \alpha^{-1})\sqrt{\beta} 
         +  {1\over 2}(1 +  \alpha )   \sqrt{\beta} ^{-1} 
         - (\sqrt{\beta} +  \sqrt{\beta} ^{-1} ) 
         = - ( 1-\alpha ) \{ 2\alpha \sqrt{\beta} (\alpha - \beta) \}^{-1}$, 
         which is negative when $\beta < \alpha < 1$ or 
         $\beta > \alpha > 1$. Thus, 
   $$
  d_k({\cal G}, {\cal G'}) + d_k({\cal G'}, {\cal G''})
    <  d_k({\cal G}, {\cal G''}) \iff
    \lambda_k < {\lambda'}_k < {\lambda''}_k \ \ {\rm or }\ \ 
    \lambda_k > {\lambda'}_k > {\lambda''}_k   \ \ \ .
    \eqno{(7)}
    $$

  Three geometries $\cal G$, $\cal G'$ and $\cal G''$ satisfying 
    $\lambda_k < {\lambda'}_k < {\lambda''}_k$ ($k=1,2,\cdots$) can 
    easily be constructed if the difference in volumes is allowed. Note 
    that the eigenvalues scale as $\lambda_k \propto V ^{-{2\over n}}$
     w.r.t.  $n$-volume $V$. Thus, any conformally equivalent 
     geometries $\cal G$, $\cal G'$ and $\cal G''$ such that 
     $vol\  {\cal G} > vol\  {\cal G'} > vol\  {\cal G''}$
      satisfy this condition.
  Thus, 
  $
  d_{\Lambda}({\cal G}, {\cal G'}) + d_{\Lambda}({\cal G'}, {\cal G''})
    <  d_{\Lambda}({\cal G}, {\cal G''})
  $ in this case. (The cut-off $\Lambda$ is needed, since volumes are 
  different.) However,  we also want to know the case of $d(\cdot, \cdot)$ 
  (corresponding to the case of $\Lambda \rightarrow \infty$).  
  We are then led to a  question: 
  {\it  Whether there exist three geometries 
  $\cal G$, $\cal G'$ and $\cal G''$ such that 
 $dim\  {\cal G} = dim\  {\cal G'} =dim\  {\cal G''}$, 
  $vol\  {\cal G} = vol\  {\cal G'} =vol\  {\cal G''}$ and 
  $\lambda_k < {\lambda'}_k < {\lambda''}_k$ ($k=1,2,\cdots$).}
  
\noindent
  This is a highly non-trivial question and it seems that  the answer is not 
  known as yet.
  
  In this connection, let us remember the distance in the superspace 
  defined by  the DeWitt metric [18]. Although the DeWitt metric
   is not the positive definite metric, the latter is induced on a 
   surface of a constant volume in the superspace. On this surface, then, 
   the distance can be defined using this positive definite metric. 
   Obviously, the distance thus 
    defined  satisfies the triangular inequality
    as  in  ordinary Riemannian geometry.
Thus, the failure of the triangular inequality for the spectral distance 
(examples of which shall  be shown 
in the next section) explicitly demonstrates that the spectral distance 
and the distance defined by the DeWitt metric are not equivalent to
each other.  
 
 \chapter{Closeness between the orientable and the non-orientable 
 universes}
 
  We   apply the spectral distance to quantum cosmology: We ask 
 a question as to whether universes with different topologies interfere
 quantum mechanically. A  probable answer which one might give would  be 
 that they decohere with each other since they `sound' differently, 
  resulting in a long spectral distance. To investigate this problem, 
 let us set up several concrete models in 2-dimension 
 with various topologies,  
 and investigate the spectral distances between them 
 in detail. In particular, we shall concentrate on the cases of 
 \item {(A)} $T^2$ and ${\bf R}P^2 \# {\bf R}P^2$ (Klein's bottle).
 \item {(B)} $S^2$ and ${\bf R}P^2$.

 Here, $T^2$ is a covering space of ${\bf R}P^2 \# {\bf R}P^2$ [11],
and the former is orientable while the latter is non-orientable. 
 The relation
 between $S^2$ and ${\bf R}P^2$ is also the same. We shall  construct models
  of $T^2$ and ${\bf R}P^2 \# {\bf R}P^2$, both of which are locally flat.
   Then, we can focus on the effect of the difference of global topologies, 
   or the difference of orientabilities, in this case. We shall also 
   construct models of $S^2$ and ${\bf R}P^2$, both of which are homogeneous
    (constant curvature spaces). By making the antipodal identification on 
    a 2-sphere with radius ${\sqrt 2} R$, a homogeneous ${\bf R}P^2$ with 
    2-volume $4\pi R^2$ can be constructed [11]. 
    Therefore, the difference between 
    $S^2$ and ${\bf R}P^2$ in this case  
    includes a  difference of  local curvature 
    as well as a difference of  orientability. 
    Thus, this is the simplest
    case in which  local as well as  global geometries are different.

 As discussed in the previous section, 
 the spectral distance can be interpreted 
 as 
$  
d({\cal G}, {\cal G'}) 
= - \ln \left\{ \rho_{reduced } ({\cal G}, {\cal G'})
  / \varphi ({\cal G})\varphi^* ({\cal G'})\right\}
$, with {\it Caveat A}. 
Thus, the spectral distances for $(A)$ and $(B)$ 
 provide us 
 with some insights for our above-mentioned question.

 \section{$T^2$ and ${\bf R}P^2 \# {\bf R}P^2$}
 
 As the simplest class of models, 
 we shall investigate spaces ($\Sigma$)  constructed 
 as $\Sigma \simeq  {\bf R}^2/G$, where $G$ 
 is a discrete subgroup of the Euclid 
 group of ${\bf R}^2$, acting freely on ${\bf R}^2$.  There are only 
 5 types of spaces constructed in this manner; 
 ${\bf R}^2$, a cylinder, a M\"obius' strip, 
 $T^2$ and ${\bf R}P^2 \# {\bf R}P^2$ [28]. Amongst  them, only 
 $T^2$ and ${\bf R}P^2 \# {\bf R}P^2$ are compact. Thus we choose these spaces 
 as our models. 
 
  {\bf 3.1.1} \ \ \  $T^2$; {\it The case of} $\tau^1 \ne 0$ \ .

A torus $T^2$ can be constructed by choosing  as $G$, 
a translation group,
$G=\{ m {\vec a} + n {\vec b} \}$ $( m,n \in {\bf Z})$, 
where ${\vec a}=(1,0)/\sqrt{\tau^2}$, 
${\vec b}=(\tau^1,\tau^2 )/\sqrt{\tau^2}$ ($\tau^2 > 0$) and 
${\bf Z}=\{0, \pm 1, \pm 2, \cdots  \}$. Here a set of two real 
parameters $(\tau^1, \tau^2)$ $(\tau^2 >0)$ are the Teichm\"uller 
parameters, characterizing the global shape of a torus [29].
\footnote\star{In this paper, $\tau^2$ always represents 
the second component of $(\tau^1, \tau^2)$, and not the square of 
$\tau:=\tau^1 + i \tau^2$. 
}
 We fix the volume of $T^2$ to be unity, which is  
 taken care of by the factor  
 $\sqrt {\tau^2}$ in the choice of ${\vec a}$ and ${\vec b}$. It is convenient 
 to introduce a coordinate $(\xi^1, \xi^2)$ defined by  
 $$
 \pmatrix{x \cr
          y \cr}= {1 \over \sqrt {\tau^2}}
               \left\{ \xi^1 \pmatrix{1 \cr 0 \cr} 
               + \xi^2 \pmatrix{\tau^1 \cr
                                \tau^2 \cr} \right\}\ \ \ .
\eqno{(8)}
 $$

In this coordinate system $(\xi^1, \xi^2)$, $\vec a$ and $\vec b$
 can be expressed as $\vec a =(1, 0)$ and $\vec b =(0, 1)$, and the 
 identification by $G$ reads
 $(\xi^1, \xi^2) \sim (\xi^1 +m, \xi^2 +n)$ for  
 $ \forall m, \forall n \in {\bf Z}$.

The flat  metric is given by 
$ds^2= dx^2 +dy^2= h_{ab} d\xi^1 d\xi^2$, where
\nextline 
$h_{ab} = {1 \over \tau^2} \pmatrix{1 & \tau^1 \cr
                                          \tau^1 & |\tau|^2 \cr}$ 
$(|\tau|^2 := (\tau^1)^2 + (\tau^2)^2)$.
Thus, the Laplacian becomes
$$
\Delta 
= {1\over {\sqrt{ h} }}\partial_a ( \sqrt{ h} h^{ab} \partial _b)
    ={1 \over \tau^2} (|\tau|^2 \partial _1^2 
    - 2 \tau^1 \partial _1 \partial _2 + \partial _2^2)\ \ \ .
$$

When $\tau^1 \ne 0$,  the spectra become,
$$
\lambda_{mn}={4\pi^2 \over \tau^2} (|\tau|^2 m^2 -2 \tau^1 mn + n^2)
          =: 4\pi^2 Q(m,n)\ \ \ , 
\eqno{(9)}
$$
with normalized eigenfunctions
$$
\cases{
\ \ \ \ u_{(0,0)}=1        & for \ \  $(m,n) =(0,0)$  \cr
 \cases{
u_{(m,n)}=\sqrt{2} \cos(2\pi m \xi^1 + 2\pi n \xi^2)  \cr
v_{(m,n)}=\sqrt{2} \sin(2\pi m \xi^1 + 2\pi n \xi^2) \cr } 
                        &{\rm for
                             \ \ $(m,n) \in {\cal R}-(0,0)$ } \ \ , 
                             \cr
}
$$
where
$
{\cal R}:={\bf N}_0\times {\bf Z} - \{ 0 \}\times (-{\bf N})
$, ${\bf N}=\{1,2, \cdots \}$ and ${\bf N}_0 :=\{0 \} \cup {\bf N}$.
It is convenient to represent the multiplicity of eigenvalues in the form 
of a `spectral diagram' as shown in {\it Figure} $1-a$.

 {\bf 3.1.2}\ \ \  $T^2$; {\it The case of } $\tau^1 = 0$\ .

The case of $\tau^1 = 0$ should be treated separately. In this case, 
the Laplacian reduces to 
$
\Delta ={1 \over \tau^2} ((\tau^2)^2 \partial _1^2 + \partial _2^2)
$. The spectra become
$$
\lambda_{mn}={4\pi^2 \over \tau^2} ((\tau^2)^2 m^2 + n^2)
          =: 4\pi^2 Q_0(m,n)\ \ ,
\eqno{(10)} 
$$
with normalized eigenfunctions
$$
\eqalign{
u_{1(m,n)}&=\cases{ 1 &  $(m,n)=(0,0)$ \cr
                    \sqrt{2} \cos 2\pi m \xi^1, \ 
                           \sqrt{2} \cos 2\pi n \xi^2  
                                       & $m,n\in {\bf N}$ \cr
                   2\cos 2\pi m \xi^1 \cos 2\pi n \xi^2
                                       & $(m,n)\in {\bf N}\times {\bf N}$ \cr
                                        } \cr
u_{2(m,n)}&=\cases{ \sqrt{2} \sin 2\pi n \xi^2 &\ \ \ \ \ \ \ \  
                               $n\in {\bf N}$\ \   $(m=0)$ \cr 
                   2\cos 2\pi m \xi^1 \sin 2\pi n \xi^2 
                                 &\ \ \  \ \ \ \ \ 
                    $(m,n)\in {\bf N}\times {\bf N}$ \cr
                                      }\cr
u_{3(m,n)}&=\cases{ \sqrt{2} \sin 2\pi m \xi^1 &\ \ \ \ \ \ \ \ 
                         $m\in {\bf N}$\ \  $(n=0) $ \cr 
                   2\sin 2\pi m \xi^1 \cos 2\pi n \xi^2 
                                 &\ \ \ \ \ \ \ \ 
                       $(m,n)\in {\bf N}\times {\bf N}$ \cr
                                      } \cr
 u_{4(m,n)} &=\matrix{ 2\sin 2\pi m \xi^1 \sin 2\pi n \xi^2  
                                 & \ \ \ \ \ \ \ \ \ \ \  
                       (m,n)\in {\bf N}\times {\bf N} \cr
                                      } \cr
}
$$
The spectral diagram for this case is  shown in {\it Figure} $1-b$.

Comparing the spectral diagram for $\tau^1 = 0$ with  the one for 
$\tau^1 \ne 0$, we see that the distribution of spectra is modified 
but  this 
modification is in such a way as to guarantee  Weyl's asymptotic 
formula for $N(\lambda)$ (Eq.$(4-b)$) to hold good. 
(By folding the diagram in {\it Figure }  $1-a$ along 
 the $m$-axis, one obtains a diagram which matches the one in 
 {\it Figure}  
 $1-b$.)

{\bf 3.1.3}\ \  ${\bf R}P^2 \# {\bf R}P^2$ {\it (Klein's bottle) }\ .

The Klein's bottle can be constructed by the point-identification 
shown in {\it Figure} 2. 

In mathematical terms, it can be constructed 
by 
${\bf R}^2 /G$ where 
$ G \simeq ({\bf Z} \times {\bf Z}) \times_{{}_{\cal S}} {\bf Z}_2$ 
($\times_{{}_{\cal S}}$: semi-direct product) [28]. 
Now, the explicit representation of 
$G$ will be given: Take $(\xi^1, \xi^2)$-space as ${\bf R}^2$. Let 
$I= diag (1,1)$, $B=diag (1, -1)$ and let 
${\vec u} = {}^t (1,0)$ and 
${\vec v} = {}^t (0,1)$ 
    ($\vec u$ and $\vec v$ are  eigenvectors of $B$ with eigenvalues 1 and  
    $-1$ respectively).
 Let $t_{\vec u}$, 
 $t_{\vec v}$ represent translations on ${\bf R}^2$ 
 by $\vec u$ and $\vec v$, respectively.
 We choose  quantities 
 of the form $(A, t_{\vec a})$ as a group element ($A=I$ or $B$) and 
  define a multiple rule as
  $
 (A, t_{\vec a}) \cdot (A', t_{\vec a'})
    = (AA', t_{\vec{ a}  + A \vec {a'} })$.    
 Then, $G$ is defined as 
  $
 G:=\{ (B, t_{\vec u})^m \cdot (I, t_{\vec v})^n | m,n \in {\bf Z} \}$. 
 By simple manipulations, 
 $$
 (B, t_{\vec u})^m \cdot (I, t_{\vec v})^n =
   (B^m, t_{m{\vec u} + (-)^m n{\vec v} } ) \ \ . 
 $$
 Here, let us note that $B^m =I$ when $m$ is even, and $=B$ when 
 $m$ is odd. 
Then, this element of $G$ acts on  any point $\vec \xi$ in ${\bf R}^2$ as 
 $
(B, t_{\vec u})^m \cdot (I, t_{\vec v})^n {\vec \xi} = 
B^m {\vec \xi} + m {\vec u} + (-)^m n {\vec v}\ \ .
$
 If we represent $(B^l, t_{m{\vec u} + n{\vec v}})$ as $(m,n : (-)^l)$ 
 ($m, n, l \in {\bf Z}$), then the multiplication rule reads, 
 $$
 (m,n : (-)^l)\cdot (m',n' : (-)^{l'}) = (m+m',n+(-)^l n' : (-)^{l+l'})
 \ \ ,
 $$                    
implying 
 $ G \simeq ({\bf Z} \times {\bf Z})\times_{{}_{\cal S}} {\bf Z}_2$.
 Furthermore, ${\bf R}^2 /G$ corresponds to the point-identification shown in 
{\it  Figure}  2, producing  Klein's bottle ${\bf R}P^2 \# {\bf R}P^2$. 

The connection of $(\xi^1, \xi^2)$ with the standard coordinate 
$(x,y)$ is the same as in Eq.(8) with the restriction $\tau^1 =0$:
Because of the particular direction of identification 
 as shown in {\it Figure} 2, the deficit
angle occurs when $\tau^1 \ne 0$, contrary to the case of $T^2$. Thus 
we shall investigate only the cases of $\tau^1 =0$. The eigenvalues become,
$$
\lambda_{mn}={4\pi^2 \over \tau^2} ((\tau^2)^2 m^2 + n^2), 
\eqno{(11-a)}
$$
with normalized eigenfunctions
$$
\matrix{
1 & \ \ \  {\rm for}\ \    (m,n)=(0,0) \cr
\sqrt{2} \cos 2\pi m \xi^1 &  {\rm  for}\ \ \    m \in {\bf N},\  n=0  \cr
\sqrt{2} \cos 2\pi n \xi^2 &  {\rm for }\ \ \   m=0,\  n \in {\bf N}  \cr
\cases{ 2\cos 2\pi m \xi^1 \cos 2\pi n \xi^2 \cr
    2\sin 2\pi m \xi^1 \cos 2\pi n \xi^2 \cr}
          &\ \ \  $for$\ \ \  m \in {\bf N},\  n \in {\bf N} \cr
}
$$
and
$$
\lambda_{m+1/2, n}={4\pi^2 \over \tau^2} ((\tau^2)^2 (m+1/2)^2 + n^2), 
\eqno{(11-b)}
$$
with normalized eigenfunctions
$$
\matrix{
\cases{2\cos 2\pi (m+1/2) \xi^1 \sin 2\pi n \xi^2 
                        \cr
       2\sin 2\pi (m+1/2) \xi^1 \sin 2\pi n \xi^2  \cr
         } & {\rm for} \ \ \ m \in {\bf N}_0,\  n \in {\bf N}\ \ . \cr
}
$$
The spectral diagram is shown in {\it Figure} 3. Compared with the diagram for 
$T^2$ ($\tau^2 =0$) ({\it Figure} $1-b$), 
the appearance of modes characterized by 
$m=$half integer is characteristic. However, the distribution of spectra 
is again in such a way as to guarantee Weyl's asymptotic
 formula, $(4-b)$, to hold good.

We should note that  the 2-volume of  Klein's bottle constructed in 
this manner is unity: We can define the integral  on a non-orientable manifold 
$\cal M$ as  half the integral on the double-covering manifold 
of  $\cal M$.
 The double-covering manifold of ${\bf R}P^2 \# {\bf R}P^2$ is $T^2$ [11]. 
 Indeed, we can see in   {\it Figure} 2 the tiles that correspond 
  to $T^2$ 
 with 2-volume being 2. (For instance, 
 a rectangle defined by points $(0, 0)$, $(2,0)$, $(2,1)$ and $(0,1)$.) 
 Thus the 2-volume of  Klein's bottle in our case  is 1.

{\bf 3.1.4}\ \ \ {\it The spectral distance between $T^2$ and 
${\bf R}P^2 \# {\bf R}P^2$ }\ \ .

 Having obtained spectra for $T^2$ and ${\bf R}P^2 \# {\bf R}P^2$
  in previous subsections, we  now proceed to 
 calculate the spectral distances for various cases according to 
 Eq.(3) or Eq.(6).
 
 First, let us see how  Weyl's asymptotic formula holds nicely 
 for checking the spectra obtained. For instance, the case of 
 $T^2$ with Teichm\"uller parameters $(\tau^1, \tau^2)=(0.1, 1)$
  is shown in {\it Figure} 4 and {\it Figure}  5.
  {\it Figure} 4 is a $\lambda_k - k$ plot. In accordance with 
  $(4-c)$ ($n=2$, $V=1$),  we see that 
  the inclination of the plot is $4\pi$. 
  {\it Figure} 5 is a $4\pi N(\Lambda)/V - \Lambda$ plot, 
  which approaches to 1 in accordance with  $(4-b)$.
  
  Next, we see the spectral distances $d(T^2, T^2)$. {\it Figure } $6-a$
   indicates  the case of $(\tau^1, \tau^2)=(0, 1)$ and 
   $(\tau^1, \tau^2)=(0, 2)$ and {\it Figure } $6-b$ shows 
   a $d_{\Lambda} ((0,1), (0,2))-\Lambda $ plot, giving  the 
   spectral distance about 0.219. The convergence of 
   the scale-dependent spectral distance $d_\Lambda$ when 
   $\Lambda \rightarrow \infty$ is fairly  good.
   
   {\it Figure } $7-a$ indicates the case of Klein's bottles with 
   $(\tau^1, \tau^2)=(0, 10)$ and $(\tau^1, \tau^2)=(0, 100)$. 
   {\it Figure } $7-b$ shows a $d_{\Lambda} ((0,10), (0,100))-\Lambda $ 
   plot. The spectral distance is about 2.916 in this case.
   
   Now, the results of calculating the spectral distances between 
   $T^2$ and
\nextline   
 ${\bf R}P^2 \# {\bf R}P^2$ are quite surprising:
   They are quite  short. For instance, 
   {\it Figure } $8-a$ indicates the case of a torus and  
   Klein's bottle with 
   $(\tau^1, \tau^2)=(0, 1)$ for a torus and $(\tau^1, \tau^2)=(0, 10)$
    for a Klein's bottle, and  
   {\it Figure } $8-b$ shows a $d_{\Lambda} -\Lambda $ 
   plot. The spectral distance is about 0.4337, which is  
   unexpectedly short.
   
{\it Figure} $9-a$, $9-b$ and $9-c$ show  some of spectral distances 
for $T^2$, Klein's bottles and mixture cases, respectively.
One can see that 
the spectral distances between $T^2$ and ${\bf R}P^2 \# {\bf R}P^2$ 
are unexpectedly short. 
A general tendency is that the spectral distance for 
$T^2-T^2$ is longer  than the one for 
$T^2 -  {\bf R}P^2 \# {\bf R}P^2$, and the same for 
${\bf R}P^2 \# {\bf R}P^2 - {\bf R}P^2 \# {\bf R}P^2 $ 
is the shortest, for fixed  
parameters $(\tau^1, \tau^2)$ and $({\tau'}^1, {\tau'}^2)$.
If we employ the density-matrix interpretation (with {\it Caveat A}),
 these results 
suggest that  an orientable universe and a non-orientable one 
sometimes 
interfere with each other quite strongly.
In other words, some extra mechanism is needed if they should 
decohere with each other.

 As has already mentioned in advance in $\S\S 2-5$, 
 many examples can be found in {\it Figure }$9-a,b,c$ 
 which do not satisfy the 
 triangular inequality,  
 $d({\cal G}, {\cal G'}) + d({\cal G'}, {\cal G''}) 
    \ge d({\cal G}, {\cal G''})$.
   For instance, in {\it Figure }$9-a$, 
   $d((0, 1),(0, 500))=68.02$,  $d((0,500),(0,1000))=57.45$, while 
   $d((0,1),(0, 1000)=137.12$; in {\it Figure }$9-c$, 
   $d((0, 1),(0, 10))=0.4337$,  $d((0,10),(0,100))=2.421$,  while 
   $d((0,1),(0, 100)=3.488$.  On the other hand, one example 
   in {\it Figure }$9-c$ which 
   satisfies the inequality is for instance, 
   $d((0, 1),(0, 50))=1.77$, 
$d((0,50),(0,100))=2.668$ and 
$d((0,1),(0, 100)=3.488$.
The failure of the   triangular inequality explicitly indicates  that 
the spectral distance and the DeWitt distance are not 
equivalent to each other.

At this stage, it is appropriate to compare the spectral distance 
and the DeWitt distance with each other more explicitly. 
 In the case of $T^2$, (the positive-definite sector of)
  the DeWitt metric reduces to the 
 Poincar\'e metric on the upper-half plane ($\tau^2 > 0$), 
 $G_{AB}={1 \over \tau^2}\  diag(1, 1)$. 
 (As is well-known, one  negative signature included in  the DeWitt metric
  corresponds to the conformal deformation in the superspace. 
  We shall  come back to this point in $\S 4$.)
  Thus, the DeWitt distance, namely the geodesic distance between 
  $(\tau^1, \tau^2)$ and $({\tau'}^1, {\tau'}^2)$ w.r.t. the 
  Poincar\'e metric is [26],[28],
  $$
 d_{DW}((\tau^1, \tau^2), ({\tau'}^1, {\tau'}^2))= 
  \ln {{1+ r} \over {1-r}}\ \ \ , 
  $$
 where $ r =\left|{ \tau-\tau'\over {\tau-{\tau'}^*} } \right| $, 
 $\tau=\tau^1 + i\tau^2$ and $\tau'={\tau'}^1 + i{\tau'}^2$.  
 In particular, 
 \nextline
 $ d_{DW}((0 , \tau^2), (0 , {\tau'}^2))
 =\left|\ln   \tau^2 / {\tau'}^2 \right|
 $, depending only on the ratio ${\tau^2 / {\tau'}^2}$. Thus, 
 for instance, 
 $d_{DW}((0 , 1), (0 , 2))= d_{DW}((0 , 10), (0 , 20))
 =d_{DW}((0 , 50), (0 , 100))=d_{DW}((0 , 500), (0 , 1000))$$=0.693$. 
 ($Figure$ 10). On the other hand, the corresponding spectral distances are 
 $d((0 , 1), (0 , 2))=0.219$, $d((0 , 10), (0 , 20))=1.14$, 
 \nextline
 $d((0 , 50), (0 , 100))=5.778$ and $d((0 , 500), (0 , 1000))=57.45$, 
 which also illustrate the non-equivalence of these two distances.

\section{$S^2$ and ${\bf R}P^2$}

The comparison of the 2-sphere with the real projective 2-space
 ${\bf R}P^2$ is another case which can be investigated with ease.
  As has already discussed at the beginning of this 
  section, one can construct the real projective space 
  ${\bf R}P^2$ of volume $4\pi R^2$  
  by the antipodal identification on a sphere with 
  radius ${\sqrt 2 } R$.  This space and a sphere with radius 
  $R$ are of same 2-volume with different orientability as our previous 
  models of tori and Klein's bottles. In the present case, however, 
  the  curvatures are also different  so that the local geometries are
   different. Thus, this case serves as the simplest case in 
   which the difference of  local geometries as well as the global 
   geometries takes part in. The spectra of $S^2$ are 
 $$
 \lambda_l = {4\pi \over V} l(l+1)\ \ \ 
 ( {\rm multiplicity}\ \  2l+1,  \ \ \  
 l=0,1,2, \cdots), 
 $$  
where $V$ is the 2-volume of $S^2$. The same of ${\bf R}P^2 $ are 
 $$
 \lambda'_l = {4\pi \over V} l(2l+1)\ \ \ 
 ( {\rm multiplicity}\ \  4l+1,  \ \ \  
 l=0,1,2, \cdots). 
 $$
 
 The spectral distance $d(S^2, {\bf R}P^2)$ again turns out to be 
 unexpectedly short about $0.8$ irrespective of the value of the 
 2-volume $V$. (See {\it Figure} 11.) 
 {\it Figure} 12 shows $d(S^2, S^2)$ with different 2-volumes. It clearly 
 shows the divergent behavior of the spectral distance when 
 volumes are not identical (see Eq.(5)). Significantly, in terms of the 
 scale-dependent spectral distance $d_\Lambda$, the degree of 
 difference in  volumes is represented by the inclination of the 
 curve of the $ d_\Lambda - \Lambda$ plot: The  larger  
  the difference in volume, the  larger the inclination 
 of the $ d_\Lambda - \Lambda$ plot. In other words, 
 although $d_\Lambda \rightarrow \infty$ as $\Lambda \rightarrow \infty$  
  when volumes are different, the asymptotic behavior 
  of $d_\Lambda$ approaching to 
  infinity still contains  the information of `closeness' between volumes.
  
  Even pure-mathematically, 
  it is interesting that some manifolds with different topology 
  (or orientability) show a very short spectral distance between 
  them as compared to  other manifolds with  identical topology.
   Furthermore, if one employs the density-matrix interpretation
   (with {\it Caveat A}) for the spectral distance, it suggests 
   that some universes with different orientabilities do not 
   decohere effectively without any other mechanisms. We shall come back 
   to this point in $\S 4$.

 \section{Epstein's zeta and theta functions,  
 and  Weyl's asymptotic formula}
 
  It  is of some interest to construct and investigate in detail 
  the heat kernel for our models, $T^2$ and 
  ${\bf R}P^2 \# {\bf R}P^2 $. In these cases, the heat kernel is 
  expressed nicely in terms of Epstein's theta functions [19],[20],[17].
   \footnote\star{
   See $Appendix$ for Epstein's theta and zeta functions.}  
  In such  cases, the functional relation for  these theta functions 
   derives Weyl's asymptotic formula,  and 
   relates directly  the expression for the heat kernel in terms of  
  mode-summation,  with  the one in terms of  image-summation [30]. 
  In particular, the case 
   of ${\bf R}P^2 \# {\bf R}P^2 $ (Klein's bottle) is non-trivial and 
   interesting as we shall see below. Furthermore, based on the 
   discussion of the analytic properties of Epstein's zeta functions as 
   meromorphic functions on ${\bf C}$, the initial condition for these  
   heat kernels can be analyzed from a different viewpoint.
 
 Thus, let us  investigate  Epstein's theta and zeta
  functions for $T^2$ and 
 \nextline 
 ${\bf R}P^2 \# {\bf R}P^2$.

{\bf 3.3.1}\ \ \  {\it The case of}\  $T^2$\ .

Let us compute the non-local zeta function,
$
\zeta (x, y:s):= \sum^{\ '}_i 
\psi_i (x) \psi_i^* (y) \lambda_i^{-s}
$, 
for $T^2$ of the case $\tau^1 \ne 0$.
The case of $\tau^1 =0$ goes almost similarly. The final result is
identical to  the result of the case $\tau^1 \ne 0$ with a
replacement of 
 $Q$ by $Q_0$ (see Eqs.(9) and (10)).
  Using results in {\bf 3.1.1} (Eq.(9) and below), 
$$
\eqalignno{
\zeta_{T^2}(\xi^{1'}\ \xi^{2'}, \xi^1\ \xi^2 : s)
 &=(4\pi^2)^{-s} Z
              {  \left\vert \left\vert \matrix{0 & 0 \cr
                 \Delta \xi^1 & \Delta \xi^2 \cr   } 
                 \right\vert \right\vert  }
  (Q,s) \cr
 &=2^{-2s} \pi^{-1} { \Gamma (1-s)\over \Gamma (s) }
        Z
          {  \left\vert  \left\vert 
          \matrix{\Delta \xi^1 & \Delta \xi^2 \cr
                 0 & 0 \cr   } 
                 \right\vert \right\vert  }
        (Q^{-1}, 1-s)\ \ \ , & (12) \cr
}
$$
where $ \Delta \xi^1 :=\xi^1 - \xi^{1'} $, 
$ \Delta \xi^2 :=\xi^2 - \xi^{2'}  $, and   
$(A5)$ in $Appendix$ has been used in the last line.

 The heat kernel 
$
K(x,y:t)= \sum _i \psi_i (x) \psi_i^* (y) \exp -\lambda_i t
$
 becomes 
$$
\eqalignno{
K_{T^2}(\xi^{1'}\ \xi^{2'}, \xi^1\ \xi^2 : t)
 &= \Theta
           {  \left\vert \left\vert \matrix{0 & 0 \cr
                 \Delta \xi^1 & \Delta \xi^2 \cr   } 
                 \right\vert \right\vert  }
  (Q,4\pi t) & (13-a) \cr
 &= {1\over 4\pi t } \Theta
          {  \left\vert  \left\vert 
          \matrix{\Delta \xi^1 & \Delta \xi^2 \cr
                 0 & 0 \cr   } 
                 \right\vert \right\vert  }
        (Q^{-1}, {1\over {4\pi t}})\ \ \ , & (13-b) \cr
}
$$
where $(A4)$  has been used in the last line.

It is clear that $(13-b)$ can be written as 
$
\sum_{m,n=-\infty}^\infty {1\over {4\pi t}}
                 \exp -Q^{-1} (m+\Delta \xi^1, n+\Delta \xi^2)/4t$, 
which is an image-summation  of the heat kernel on ${\bf R}^2$. 
Thus, the functional relation $(A4)$ guarantees the equivalence between 
 mode-summation $(13-a)$ and image-summation $(13-b)$ [30].

Clearly,  
$K_{T^2}$ satisfies the heat equation, 
$
\Delta _{\vec \xi} K (\vec \xi, \vec \xi':t)
= \partial /\partial t K (\vec \xi, \vec \xi':t)
$.
 It is of some interest to clarify the initial condition which 
 $K_{T^2}$ should satisfy:
From Eq. $(13-b)$, 
$$
\eqalign{
\lim_{t \downarrow 0} K_{T^2} (\vec \xi, \vec \xi':t)
 = & \lim_{t \downarrow 0} 
   {1\over {4\pi t}} 
   \exp -Q^{-1} (\Delta \xi^1, \Delta \xi^2)/4t \cr
   & + \lim_{t \downarrow 0} \sum_{m,n}{{}'} 
   {1\over {4\pi t}} \exp -Q^{-1}(m+\Delta \xi^1, n+ \Delta \xi^2)/4t
   \ \ \ . \cr
}
$$
The first term is a local contribution while the second term is a 
non-local one coming from the point-identification. The first term 
is equivalent to $\delta (\Delta \xi^1) \delta (\Delta \xi^2)$. 
 The second term can be written as 
 $$
\eqalign{
\lim_{t \downarrow 0} & \sum _{m,n}{{}'}
  {1\over {4\pi t}} \sum_{k=0}^{\infty} 
         {(-)^k \over k! } \left[ Q^{-1} 
        (m+\Delta \xi^1, n+\Delta \xi^2 )\right] ^k {1 \over (4t)^k} \cr
& = \lim_{t \downarrow 0} 
  {1\over {4\pi t}} \sum _{k=0}^{\infty} {(-)^k \over k! }
    {1 \over (4t)^k} Z
           {  \left\vert \left\vert 
              \matrix{ \Delta \xi^1 & \Delta \xi^2 \cr
                     0 & 0 } \right\vert \right\vert  }
                     (Q^{-1}, -k)\ \ \ . \cr
}
$$    
This should vanish because 
$
             Z
              {  \left\vert \left\vert \matrix{ {\vec g} \cr
                     {\vec h} } \right\vert \right\vert  }
  (Q,s)  
$
 (as a meromorphic function extended onto $\bf C$) has 
 simple zeros at $s=-1,-2, \cdots$, and furthermore, $s=0$ is also a 
 simple zero when ${\vec g} \notin {\bf Z}^N$
  (see $Appendix$). Thus $K_{T^2}$ satisfies 
 the ordinary initial condition for a heat-kernel. 
 It may be noted  that if the 
 limit $\lim _{\Delta \xi ^{1,2} \rightarrow 0}$ is taken before 
 $\lim _{t \downarrow 0}$, the second term behaves as 
 $\lim _{t \downarrow 0} {1\over 4 \pi t} \sim \delta (0)$. Thus, 
 $\lim _{\Delta \xi ^{1,2} \rightarrow 0} \lim _{t \downarrow 0}
  \ne \lim _{t \downarrow 0} \lim _{\Delta \xi ^{1,2} \rightarrow 0}$ 
   as for the second term,  because of the special zero-structure 
   of the zeta function.

 From Eq.$(13-b)$, one  can easily derive  Weyl's asymptotic formula for 
 \nextline
 $\sum_i \exp -\lambda_i t$ (see $(4-a)$),
$$
\eqalignno{
K_{T^2}(t) &= Tr K_{T^2} (\vec \xi, \vec \xi':t)       
           = {1 \over 4\pi t} \Theta 
                     {  \left\vert \left\vert 
                       \matrix{ 0 & 0 \cr
                           0 & 0 } \right\vert \right\vert  }
                         (Q^{-1}, {1 \over 4\pi t}) \cr
           &= {1 \over 4\pi t} \sum_{m,n} \exp -\pi Q^{-1}(m,n)/4\pi t 
     = {1 \over 4\pi t} \sum_{k=0}^\infty
          { (-)^k \over {(4t)^k k!} } Z 
                     {  \left\vert \left\vert 
                       \matrix{ 0 & 0 \cr
                           0 & 0 } \right\vert \right\vert  }
                       (Q^{-1},-k) \cr
     &= {1 \over 4\pi t}\ \ \ , &(14) \cr            
}
$$
where the zero-structure  of 
$Z 
                     {  \left\vert \left\vert 
                       \matrix{ 0 & 0 \cr
                           0 & 0 } \right\vert \right\vert  } $
has again been used. Noting $(4-a)$ and below, this result matches 
the fact that $V=1$,  
 $\chi (T^2) =0$.

{\bf 3.3.2} \ \ \ 
{ \it The case of } ${\bf R}P^2 \# {\bf R}P^2$ {\it (Klein's bottle)}\ .

We can proceed in an almost  parallel manner as in the case of 
$T^2$. The non-local zeta function becomes,
$$
\eqalignno{
\zeta_{Klein}&(\xi^{1'}\ \xi^{2'},  \xi^1\ \xi^2 : s) \cr 
 &=  {1 \over 2}(4\pi^2)^{-s} \big\{
    Z
                \left\vert \left\vert \matrix{0 & 0 \cr
                 \Delta \xi^1 & \Delta \xi^2 \cr   } 
                 \right\vert \right\vert  
  (Q_0,s)
    +
    \cos\pi \Delta \xi^1 \ 
    Z
                \left\vert \left\vert \matrix{1/2 & 0 \cr
                 \Delta \xi^1 & \Delta \xi^2 \cr   } 
                 \right\vert \right\vert  
  (Q_0,s)                \cr  
      &   + Z
                \left\vert \left\vert \matrix{0 & 0 \cr
                 \Delta \xi^1 & \Delta_+ \xi^2 \cr   } 
                 \right\vert \right\vert  
  (Q_0,s)
   - \cos\pi \Delta \xi^1 \  
   Z
                \left\vert \left\vert \matrix{1/2 & 0 \cr
                 \Delta \xi^1 & \Delta_+ \xi^2 \cr   } 
                 \right\vert \right\vert  
  (Q_0,s)     \big\}\ , &(15)
    \cr
  }
$$
where $\Delta \xi^i := \xi^i - \xi^{i'}$ ($i=1,2$), 
$ \Delta_+ \xi^2 := \xi^2 + \xi^{2'} $. Taking the trace of this 
expression, the local zeta function becomes,
$$
\eqalignno{
\zeta_{Klein}( s)  
 &=  {1 \over 2}(4\pi^2)^{-s} \big\{
    Z
                \left\vert \left\vert \matrix{0 & 0 \cr
                 0 & 0 \cr   } 
                 \right\vert \right\vert  
  (Q_0,s)
    +
    Z
                \left\vert \left\vert \matrix{1/2 & 0 \cr
                 0 & 0 \cr   } 
                 \right\vert \right\vert  
  (Q_0,s)                \cr  
      &  \ \ \ \ \ \ + Z
                \left\vert \left\vert \matrix{0  \cr
                 0  \cr   } 
                 \right\vert \right\vert  
  (Q_0(\cdot ,0),s)
   -  Z
                \left\vert \left\vert \matrix{1/2  \cr
                 0 \cr   } 
                 \right\vert \right\vert  
  (Q_0 (\cdot ,0),s)     \big\}\ \ \ .  & (16) \cr
  }
$$
 The first terms in (15) and (16) are similar to the case of $T^2$, 
 while the second terms originate from the half-integer modes 
 $\lambda_{m+1/2, n}$, characteristic for   the case of  Klein's
  bottle. The last two terms  in (15) and (16) reflect the spatial
   inhomogeneity of the present  model. 
   ( $\Delta_+ \xi^2 \rightarrow 2 \xi^2$ when 
   $  \Delta \xi^1, \Delta \xi^2 \rightarrow  0  $, indicating 
   the spatial dependence.)

The heat-kernel for  Klein's bottle becomes,
$$
\eqalignno{
K&_{Klein}(\xi^{1'}\ \xi^{2'},  \xi^1\ \xi^2 : t) \cr
& =  {1 \over 2}
    \Theta
                \left\vert \left\vert \matrix{0 & 0 \cr
                 \Delta \xi^1 & \Delta \xi^2 \cr   } 
                 \right\vert \right\vert  
  (Q_0,4\pi t) 
    +
    {1 \over 2}\cos\pi \Delta \xi^1 
    \Theta
                \left\vert \left\vert \matrix{1/2 & 0 \cr
                 \Delta \xi^1 & \Delta \xi^2 \cr   } 
                 \right\vert \right\vert  
  (Q_0,4\pi t)                 \cr 
         & + {1\over 2} \Theta 
                \left\vert \left\vert \matrix{0 & 0 \cr
                 \Delta \xi^1 & \Delta_+ \xi^2 \cr   } 
                 \right\vert \right\vert  
  (Q_0, 4\pi t)
   - {1\over 2} \cos\pi \Delta \xi^1 
   \Theta
                \left\vert \left\vert \matrix{1/2 & 0 \cr
                 \Delta \xi^1 & \Delta_+ \xi^2 \cr   } 
                 \right\vert \right\vert  
  (Q_0,4\pi t)\ \ . \cr 
  &  &(17-a) \cr
 } 
$$

The application of the functional relation $(A4)$ to the heat-kernel
\nextline
 $K_{Klein}(\xi,  \xi' : t)$, $(17-a)$, which has been obtained
  by the mode-summation, provides 
 a non-trivial expression for  the heat-kernel corresponding to 
 the point-summation:
 $$
 \eqalignno{
&K_{Klein}(\xi^{1'}\ \xi^{2'},  \xi^1\ \xi^2 : t) \cr
& =  {1 \over 2}\cdot {1 \over 4 \pi t} 
     \big\{
    \Theta
                \left\vert \left\vert \matrix{
                 \Delta \xi^1 & \Delta \xi^2 \cr
                 0 & 0 \cr   } 
                 \right\vert \right\vert  
  (Q_0^{-1},{1 \over 4\pi t}) 
    + \cos\pi \Delta \xi^1  
    \Theta
                \left\vert \left\vert \matrix{
                 \Delta \xi^1 & \Delta \xi^2 \cr
                 1/2 & 0 \cr   } 
                 \right\vert \right\vert  
         (Q_0^{-1},{1 \over 4\pi t})          \cr 
        & + 
         \Theta
                \left\vert \left\vert \matrix{
                 \Delta \xi^1 & \Delta_+ \xi^2 \cr 
                 0 & 0 \cr  } 
                 \right\vert \right\vert  
  (Q_0^{-1},{1 \over 4\pi t})
   -  \cos\pi \Delta \xi^1 
   \Theta
                \left\vert \left\vert \matrix{
                 \Delta \xi^1 & \Delta_+ \xi^2 \cr
                 1/2 & 0 \cr   } 
                 \right\vert \right\vert  
  (Q_0^{-1},{1 \over 4\pi t})
  \big\}
  \ \ \  . \cr
  & & (17-b) \cr
 } 
 $$
This reduces to the expression 
$$
\eqalignno{
K_{Klein}&(\xi, \xi':  t) \cr
&= \sum_{m,n=-\infty}^{\infty} \cos^2 {\pi \over 2} (m + \Delta \xi^1)\ 
       {1 \over 4\pi t}\  \exp -Q_0^{-1}
       (m + \Delta \xi^1, n + \Delta \xi^2)/4t \cr
     & 
   +\sum_{m,n=-\infty}^{\infty} \sin^2 {\pi \over 2} (m + \Delta \xi^1)\ 
       {1 \over 4\pi t}\  \exp -Q_0^{-1}
       (m + \Delta \xi^1, n + \Delta_+ \xi^2)/4t \ ,\cr
& & (17-b') \cr
}
$$ 
which is in the form of a point-summation of the heat-kernel 
for ${\bf R}^2$ in a non-trivial manner. When $\Delta \xi^1 =0$, 
only the terms for $m=$even (odd) remain in the first (second) 
summation. Then the first term matches $K_{T^2}$ for a torus 
constructed from $1 \times 2$ band in $(\xi^1, \xi^2)$-space, 
which corresponds to the  covering space [11] of our 
Klein's bottle. The second term originates from the twisting in the 
point-identification for constructing the model.

Taking the trace of this, the corresponding theta function becomes,
$$
\eqalign{
K_{Klein} (t) &:=\sum_i \exp -\lambda_i t \cr
 &=  {1 \over 2}
    \Theta
                \left\vert \left\vert \matrix{0 & 0 \cr
                 0 & 0 \cr   } 
                 \right\vert \right\vert  
  (Q_0,4\pi t)
    +
    {1 \over 2} \Theta
               \left\vert \left\vert \matrix{1/2 & 0 \cr
                 0 & 0 \cr   } 
                 \right\vert \right\vert  
  (Q_0, 4\pi t)                \cr  
      &  \ \ \ \ \ \ + {1 \over 2} \Theta
                \left\vert \left\vert \matrix{0  \cr
                 0  \cr   } 
                 \right\vert \right\vert  
  (Q_0(\cdot ,0), 4\pi t)
   -  {1 \over 2} \Theta
                \left\vert \left\vert \matrix{1/2  \cr
                 0 \cr   } 
                 \right\vert \right\vert  
  (Q_0 (\cdot ,0), 4\pi t )\ \ \ . \cr
 }
 $$
The interpretation of each term is similar to  the case of 
 the zeta function.

The functional relation helps us to get further insight. By the use of
 $(A4)$, $K_{Klein} (t) $ becomes 
$$
\eqalignno{
K_{Klein}& (t) 
 =  {1 \over 2}\cdot {1 \over 4\pi t}
 \big\{
    \Theta
                \left\vert \left\vert \matrix{0 & 0 \cr
                 0 & 0 \cr   } 
                 \right\vert \right\vert  
  (Q_0^{-1} ,{1 \over 4\pi t})
    +
     \Theta
               \left\vert \left\vert \matrix{0 & 0 \cr
                 1/2 & 0 \cr   } 
                 \right\vert \right\vert  
  (Q_0^{-1},{1 \over  4\pi t})                \cr  
        +  & {1 \over \sqrt{\tau^2}} 
        \Theta
                \left\vert \left\vert \matrix{0  \cr
                 0  \cr   } 
                 \right\vert \right\vert  
  (Q_0(\cdot ,0)^{-1}, {1 \over 4\pi t})
   -   {1 \over \sqrt{\tau^2}}
 \Theta
                \left\vert \left\vert \matrix{0  \cr
                 1/2 \cr   } 
                 \right\vert \right\vert  
  (Q_0 (\cdot ,0)^{-1}, { 1 \over 4\pi t} ) \big\}  \cr
  &={1 \over 4 \pi t}  \ \ \ , & (18) \cr
}
$$
where we have followed the same discussion as the case of 
$K_{T^2} (t)$. This result coincides with  Weyl's 
asymptotic formula $(4-a)$ corresponding to our model 
($n =2$, $V =1$, Euler number$=0$).

\chapter{Discussions}

Let us now compare the spectral distance and the DeWitt distance with 
each other. 
There are several differences in between them.

  First,
 the DeWitt metric appears within the realm of general relativity, 
 while the spectral distance has been introduced from a general 
 argument ($\S\S 2-1$), which is itself  independent of general relativity.
 
  Secondly, the DeWitt metric is a metric which can be read out from 
 the structure of the `kinetic term' of the Hamiltonian constraint of 
 general relativity. Just as  the geometrical structure of 
 the configuration space of an ordinary mechanical system 
 reflects on the kinetic term, the DeWitt metric reflects the 
 geometrical structure of the superspace. On the other hand, 
 the spectral distance is the measure of the difference in `sounds' of 
 two universes. In other words, a suitable matter field is used 
 as a probe of local and global geometry of the universe. In the 
 above sense, the DeWitt distance may be called as 
 a `kinematical distance', 
 while the spectral distance  may be called as a `dynamical distance'.
 
 Next, by construction, 
 the DeWitt distance can be defined only between universes in  
 the same superspace. The spectral distance can be 
 defined in principle between any kind of universes. As has been discussed 
 in $\S\S 2-3$  and $\S\S 2-4$, the spectral distance requires 
 a cut-off $\Lambda$  for some
  cases, like in the case of universes with different dimension or 
  different volume. Significantly, even then, the asymptotic behavior 
  of $d_\Lambda$ as $\Lambda \rightarrow \infty$ still contains 
  information of the `closeness' between two universes as has been 
  discussed in $\S\S 3-2$. 
  
  The most striking difference is that 
  the triangular inequality holds for the DeWitt distance by 
  construction, while  it fails to hold in general for the 
  spectral distance ($\S\S 2-5$). (Rigorously 
  speaking, therefore, $d({\cal G}, {\cal G'})$ should be called 
  as like a `measure of closeness', and not a `distance'.)
   This fact explicitly 
  demonstrates the non-equivalence between 
  two distances. There is still a possibility of choosing 
  the function $F(x)$ suitably to make $d({\cal G}, {\cal G'})$ 
  satisfy the triangular inequality ($\S\S 2-1$). Even though the 
  choice of $F(x)=\exp -x$ causes the failure of 
  the triangular inequality, however, it is still distinguished from 
  other possible choices since it derives the density matrix 
  interpretation of the spectral distance ($\S\S 2-2$).     

 We have enumerated  the differences  between the two kinds of distances
  above. 
 However, we can also draw some interesting parallels 
 between them.  
 Let ${\cal G}$ and ${\cal G'}$ are of the same dimension and topology.
 
 As far as the models investigated in $\S 3$ concerned, 
  both distances well correlate   our intuitive notion of `similar (or 
  different) shapes'.  Although the two distances are not equivalent to 
  each other, correlations between them are quite strong.  

Another interesting parallelism appears when  volumes are different.  
The DeWitt metric is not a positive definite metric, but it includes 
one negative signature. This negative signature corresponds to 
the direction of conformal deformation in the superspace (i.e. the 
change of volume preserving information of angles). If two geometries 
$\cal G$ and $\cal G'$ lie on different surfaces of constant volume in 
the superspace, therefore, the DeWitt distance cannot be defined between 
them. Similarly, the spectral distance between $\cal G$ and 
$\cal G'$ becomes divergent in this case ($\S\S 2-3$). Significantly, 
the asymptotic behavior of the scale-dependent spectral distance 
$d_\Lambda (\cal G, \cal G') $ as $\Lambda \rightarrow \infty$ 
still provides the information of the `closeness' of volumes ($\S\S 3-2$).
The above observation suggests that the difference in volume seems to be  
quite different in nature from the other differences in geometry. 
Combined with the density matrix interpretation, it suggests that 
the difference in volume causes a very strong decoherence between two 
universes ($\S\S 2-2$). This observation also suggests the modified way 
of comparing two geometries: Separating the information of volume 
and the conformal geometry as  ${\cal G}= (V, \tilde {\cal  G}) $ 
($vol\  {\tilde { \cal  G}} = 1$), a set  $(V,V') $  and 
$d({\tilde {\cal  G}}, {\tilde {\cal  G'}})$ may be used as a measure 
of closeness between $\cal G$ and $\cal G'$.

Now, let us discuss about the spectral distance between universes 
with 
different topologies. To extract the pure topological effects, we have 
prepared models of $T^2$ and Klein's bottles: Both are locally flat
and $T^2$ corresponds to 
 a double-covering space of  Klein's bottle. Thus, they are
 locally of the same geometry,  and 
the difference between them is purely topological. 
We have then investigated $d(T^2, T^2)$,
 $d({\bf R}P^2 \# {\bf R}P^2, {\bf R}P^2 \# {\bf R}P^2)$ and 
 $d(T^2, {\bf R}P^2 \# {\bf R}P^2)$ with various 
 Teichm\"uller parameters. As a result, 
$d(T^2, {\bf R}P^2 \# {\bf R}P^2)$ has turned out to be quite  
short compared with $d(T^2, T^2)$ and 
 $d({\bf R}P^2 \# {\bf R}P^2, {\bf R}P^2 \# {\bf R}P^2)$, 
 taking into account that $T^2$ and ${\bf R}P^2 \# {\bf R}P^2$ 
 are topologically different ($\S\S 3.1.4$). 
 
 Furthermore, 
 we have also investigated the cases of 
 $S^2$ and ${\bf R}P^2$ (both are homogeneous and of the same 
 2-volume). This time, $S^2$ is again a double-covering 
 space of ${\bf R}P^2$, so that the difference in topology 
 is minimal. In addition to this, local geometries are also different 
 in this case, 
 since  (constant)  scalar curvatures  are different though both of them are 
 homogeneous. (In this sense, the difference in local geometry is also 
 minimal.) As a result, $d(S^2, {\bf R}P^2 )$ has again turned out 
 to be relatively short ($\S\S 3-2$).
 
 Now, we come back to the original question posed at the beginning 
 of $\S 3$: Whether universes with different topologies interfere
 quantum mechanically? A  probable answer presented there has been  
 that they decohere with each other strongly since they 
 `sound' differently, 
  resulting in a long spectral distance. However, 
  the above results  
  suggest that this answer is not enough to 
  explain everything. We now know that there are at least 
  some cases in which 
  the spectral distance between 
   two spaces with different orientabilities 
  becomes  very short.
  
 Clearly, further investigations are needed 
  to  clarify this point. We need to investigate to what extent 
  it is of generality that 
  the spectral distance becomes very short 
  between two spaces $\cal G$ and $\cal G'$, where  
  ${\cal G} = (\Sigma , g)$,   ${\cal G'} = (\Sigma / G , g)$, and  
  $G$ is a discrete subgroup of the isometry group of $(\Sigma , g)$ 
  (like our models of $T^2$ and Klein's bottles). We also need 
  to investigate whether the spectral distance between two spaces 
  with more drastic difference in topology becomes large. For instance, 
  the case of two hyperbolic surfaces with a different genus should be 
  investigated.  In this case,  the properties of spectra ( and 
  the `length spectra'  (a set of lengths of all elementary closed 
  geodesics), which are in some sense the dual concept of the spectra)
   are extensively investigated by means of Selberg's trace formula 
   [26],[31]. 
   At the same time,  in this case, numerical methods are also required 
   to get explicit spectra. 
    This case of hyperbolic surfaces 
   may be  an appropriate case as the next step of 
  investigations.
 
Finally, it is appropriate to mention the relation of the 
spectral representation with the index theorems [32]. 
They have some similarity 
in the sense that both of them connect the eigenvalues of some 
elliptic operator on a space,  with the  topological structures of the 
 space. It is clear, however, that the spectral representation
  provides a finer measure than the index theorems. 
  This can be seen in many respects.
 For instance, the index theorems talk about the analytical index 
 $ind D$, which is characterized  by the zero modes, 
 $ind D:=\sum_{j=0}^m (-)^j dim\  ker \Delta_j$, or for the simplest case,
 $= dim\  ker D - dim\  ker D^\dagger$ [32]. On the other hand, the 
 spectral representation looks at the whole spectra. This $ind D$ takes 
 the value in $\bf Z$, while the spectral distance varies in $\bf R$.
 As is seen in the examples of \S 3, for  flat tori, 
 the spectral distance even senses the difference 
 in the Teichm\"uller parameters $(\tau^1, \tau^2)$. 
 On the other hand, we can draw  many parallels between  
 the discussions of the spectral distance, and those of  
 the index theorems in the context of  the anomalies 
 of  gauge theories [32]. It is 
 interesting to investigate to what extent these two concepts are
understood  
 in a unified picture.

\ack{
The author wishes to thank M. Sakagami and S. Jhingan  
for valuable discussions and encouragements. This work is 
a part of the Indo-Japanese Cooperative
Scientific Research Programme  by  the Japan Society for the 
Promotion of Science. This work has also been  supported by 
the Yukawa Memorial Foundation, the Japan Association for Mathematical 
Sciences,  and the Japan Society for the 
Promotion of Science.}

\appendix

Here, we shall briefly discuss the basic properties of the 
theta function and the zeta function due to  Epstein. 

Let $Q$ be a $N \times N$   symmetric positive definite matrix 
(so that $det \ Q >0$). For brevity, let us denote the quadratic 
form defined by $Q$ as 
$Q(x_1, x_2, \cdots, x_N) = Q({\vec x})$
\nextline
$  := (x_1, x_2, \cdots, x_N )\  Q\ {}^t (x_1, x_2, \cdots, x_N ) $. 
  The Epstein's theta function and zeta function are defined as [19], 
  respectively, 
 $$
 \eqalignno{
\Theta   
                \left\vert \left\vert \matrix{g_1 \cdots g_N \cr
                 h_1 \cdots h_N \cr   } 
                 \right\vert \right\vert  
  (Q,t)
  := &
  \sum_{n_1, \cdots , n_N = -\infty} ^\infty
   \exp 2\pi i (n_1 h_1+ n_2 h_2 + \cdots + n_N h_N) \times \cr
   & \ \ \  
        \times  \exp -\pi Q (n_1+g_1, n_2+g_2, \cdots , n_N+g_N)\  t\  , 
                          & (A1) \cr
   Z 
                \left\vert \left\vert \matrix{g_1 \cdots g_N \cr
                 h_1 \cdots h_N \cr   } 
                 \right\vert \right\vert  
  (Q,s)
  :=& 
  \sum_{n_1, \cdots , n_N = -\infty} ^{\infty_{\ {}'} } 
   \exp 2\pi i (n_1 h_1+ n_2 h_2 + \cdots + n_N h_N) \times \cr
 & \ \ \ \ \ \  \times 
 \left(  Q(n_1+g_1, n_2+g_2, \cdots , n_N+g_N ) \right)^{-s}\ \ , 
 & (A2) \cr
 ( {\rm For} \  Re\ s > N/2,\ &
   {\rm with}\  {\rm the}\ {\rm analytic}\  
   {\rm continuation}\  {\rm onto}\  {\bf C}) \cr
 }
$$
where the prime on the sigma-symbol in $(A2)$ indicates that
 $(n_1,  n_2, \cdots , n_N ) =$
 \nextline
 $ - (g_1,  g_2, \cdots , g_N )$ 
 should be excluded from the summation  
 when $(g_1,  g_2, \cdots , g_N ) \in {\bf Z}^N$,  
 to avoid divergence. 
 Introducing  
 the  vector-notation, $(A1)$ and $(A2)$ can be expressed as, 
 respectively,  
  $$
  \eqalignno{
 \Theta   
                \left\vert \left\vert \matrix{{\vec g} \cr
                                            {\vec h} \cr   } 
                 \right\vert \right\vert  
  (Q,t)
  := &
  \sum_{ {\vec n} \in {\bf Z}^N }
   \exp 2\pi i { \vec n}\cdot {\vec h} \ 
   \exp -\pi Q ({\vec n} + {\vec g})\  t\ \ , & (A1') \cr
 Z   
                \left\vert \left\vert \matrix{{\vec g} \cr
                                            {\vec h} \cr   } 
                 \right\vert \right\vert  
  (Q,s)
  := & 
  \sum_{ {\vec n} \in {\bf Z}^N }{}'
   \exp 2\pi i { \vec n}\cdot {\vec h} \ \ 
   Q({\vec n} + {\vec g})^{-s} \ \ . & (A2') \cr
 }
 $$

 These are related to each other 
by the Mellin transformation, 
$$
{\bf M}_s \left(
 \Theta   
                \left\vert \left\vert \matrix{{\vec g} \cr
                                            {\vec h} \cr   } 
                 \right\vert \right\vert  
  (Q, \cdot) 
  - \delta_{\vec g}\  \exp -i2\pi {\vec g}\cdot {\vec h} \right)
= \pi^{-s} \Gamma (s) 
   Z   
                \left\vert \left\vert \matrix{{\vec g} \cr
                                            {\vec h} \cr   } 
                 \right\vert \right\vert  
  (Q,s)\ \ \ ,
  \eqno{(A3)}
$$
where,  ${\bf M}_s f(\cdot) = \int _0 ^\infty dt \ t^{s-1} f(t)$, and 
 $\delta_{\vec g}=1$ when ${\vec g} \in {\bf Z}^N$,  $=0$ otherwise.

They are generalizations of  Jacobi's theta function and  Riemann's 
zeta function. (More general definitions than $(A1)$ 
and $(A2)$ are possible [20], but the above are sufficient 
for our purposes.) Just like Jacobi's theta function and  Riemann's 
zeta function, they satisfy the functional relations, which are
 very useful for physical applications [30]:
 $$
 \Theta
 { \left\vert \left\vert \matrix{ {\vec g} \cr
                     {\vec h} } \right\vert \right\vert }
 (Q,t)
 = {\exp { -i2\pi {\vec g}\cdot {\vec h} } \over 
      { \sqrt{det Q} \  t^{N/2}   }  } \cdot 
 \Theta
 { \left\vert \left\vert \matrix{ -{\vec h} \cr
                     {\vec g} } \right\vert \right\vert }
 (Q^{-1},{1 \over t})\ \ \ ,
 \eqno{(A4)}
  $$
  $$
\pi ^{-s} \Gamma (s) 
             Z
              {  \left\vert \left\vert \matrix{ {\vec g} \cr
                     {\vec h} } \right\vert \right\vert  }
  (Q,s)   
=
\pi ^{-({N\over 2} -s)} \Gamma ({N \over 2} -s) \cdot
{   \exp { -i2\pi {\vec g}\cdot {\vec h} }  \over 
    \sqrt{det Q}  } \cdot
             Z
              {  \left\vert \left\vert \matrix{ -{\vec h} \cr
                     {\vec g} } \right\vert \right\vert  }
  (Q^{-1}, {N\over 2}-s).
  \eqno{(A5)} 
$$

The analytic continuation  makes the zeta function expressed as 
in $(A2)$ a meromorphic function on $\bf C$. Its pole-structure 
and  zero-structure as a meromorphic function are as follows:
\item{(i)}
When ${\vec h} \notin {\bf Z}^N$, it is holomorphic on $\bf C$. 
 There are simple 
zeros at least at $s= -1,-2, \cdots$. Furthermore, there occurs 
one more simple zero at 
$s=0$ {\it iff} ${\vec g} \notin {\bf Z}^N $.
\item{(ii)}
When ${\vec h} \in {\bf Z}^N$, 
 there is  a simple pole at $s=N/2$ with  residue  
 ${\pi ^{N/2} \over \Gamma (N/2) }  
 {1 \over \sqrt{ det\ Q}}$. 
Simple 
zeros at least at $s= -1,-2, \cdots$. Furthermore, there occurs 
one more simple zero at 
$s=0$ {\it iff} ${\vec g} \notin {\bf Z}^N $.

Note that the above-mentioned zeros are only the ones which are 
found from the  discussions of analytic properties (`trivial zeros'). 
Nothing definite can be said for other zeros even for the simplest case, 
i.e. the case of Riemann's zeta function (corresponding to the case of 
$Q \equiv 1$, ${\vec g} = {\vec h}={\vec 0}$) 
(one may remember the `Riemann conjecture' [33]).

  We now prove $(A4)$. Let us remember the Poisson's summation formula [34] ,
$$
\sum_{{\vec n} \in {\bf Z}^N} \psi ({\vec h} + {\vec n} )
 = \sum_{{\vec n} \in {\bf Z}^N} 
   \exp i 2\pi {\vec n}\cdot {\vec h} \  { \hat \psi} ({\vec n})\ \ \ ,
\eqno{(A6)}
$$  
where ${ \hat \psi}$ is the Fourier transformation of $\psi$:
$
{ \hat \psi} ({\vec k}) = \int_{-\infty}^{\infty} \psi ({\vec x})
 \ \exp - i 2\pi  {\vec k} \cdot {\vec x}\  d{\vec x}
 $, 
 $
 \psi ({\vec x}) = \int_{-\infty}^{\infty} {\hat \psi} ({\vec k})
 \ \exp i 2\pi  {\vec k} \cdot {\vec x}\  d{\vec k}
 $. (Putting the $2 \pi$ in the exponential is just for 
 the neatness of the formulas [34].)
Taking ${\hat \psi}({\vec k})
= \exp -\pi Q({\vec k}+ {\vec g})\ t $, the R.H.S. of $(A6)$ becomes
 $\Theta
 { \left\vert \left\vert \matrix{ {\vec g} \cr
                     {\vec h} } \right\vert \right\vert }
 (Q,t)
$, which in turn is the L.H.S. of $(A4)$.
 For this choice of ${ \hat \psi}$, its  inverse Fourier transformation 
 becomes, 
 $$
 \psi ({\vec x})= 
 {\exp { -i2\pi {\vec g}\cdot {\vec x} } \over 
      { \sqrt{det Q} \  t^{N/2}   }  }
      \exp -\pi Q^{-1} ( {\vec x}) {1 \over t}\ \ \ .
 $$  
Thus, the L.H.S. of $(A6)$ turns out to be the R.H.S. of $(A4)$, 
which proves $(A4)$.

Noting $(A3)$, 
one basically performs the Mellin transformation of 
the both sides of $(A4)$
 to derive $(A5)$, but a bit of care should be taken for the case of 
 ${\vec g} \in {\bf Z}^N $. 
 Thus, let us  set
 $ \phi
 { \left\vert \left\vert \matrix{ {\vec g} \cr
                     {\vec h} } \right\vert \right\vert }
 (Q,t)
= 
\Theta
 { \left\vert \left\vert \matrix{ {\vec g} \cr
                     {\vec h} } \right\vert \right\vert }
 (Q,t) 
 - \delta_{\vec g} \exp -i2\pi {\vec g}\cdot {\vec h}$. 
  Then,
 $$
 \eqalignno{
 \phi
 { \left\vert \left\vert \matrix{ {\vec g} \cr
                     {\vec h} } \right\vert \right\vert }
 (Q,t)
 & =
 -\delta_{\vec g} \exp -i2\pi {\vec g}\cdot {\vec h}
 + {1 \over \sqrt{det \ Q}\  t^{ N \over 2}}\delta_{\vec h} \cr
  & + {1 \over \sqrt{det \ Q}\  t^{ N \over 2}}  
 \exp -i2\pi {\vec g}\cdot {\vec h} \ 
 \phi
 { \left\vert \left\vert \matrix{ {-\vec h} \cr
                     {\vec g} } \right\vert \right\vert }
 (Q^{-1},t^{-1}) & (A7) \cr
}
 $$
 because of $(A4)$.
Then, its  Mellin transformation becomes
$$
\eqalignno{
{\bf M}_s
 \phi &
 { \left\vert \left\vert \matrix{ {\vec g} \cr
                     {\vec h} } \right\vert \right\vert }
 (Q,\cdot)
 =\left( \int_0^1 + \int_1^{\infty} \right)
    \phi
   { \left\vert \left\vert \matrix{ {\vec g} \cr
                     {\vec h} } \right\vert \right\vert }
   (Q,t)t^{s-1}dt \cr
& =\int_1^\infty
    \phi
   { \left\vert \left\vert \matrix{ {\vec g} \cr
                     {\vec h} } \right\vert \right\vert }
   (Q,1/t)t^{-1-s}dt
+\int_1^\infty
    \phi
   { \left\vert \left\vert \matrix{ {\vec g} \cr
                     {\vec h} } \right\vert \right\vert }
   (Q,t)t^{s-1}dt \cr
& =
{ \exp -i2\pi {\vec g}\cdot{\vec h} \over \sqrt{det Q} }
\big\{
-{ {\exp -i2\pi (-{\vec h})\cdot {\vec g} } \over {N/2 -s} }
                 \delta_{-{\vec h} }
- {1 \over \sqrt{det Q^{-1} }} {1 \over (N/2 -s)-N/2}
           \delta_{\vec g} \cr
&\ \ + { \exp -i2\pi ( -{\vec h})\cdot{\vec g}  \over \sqrt{det Q^{-1} }} 
\int_1^\infty
    \phi
   { \left\vert \left\vert \matrix{ {\vec g} \cr
                     {\vec h} } \right\vert \right\vert }
   (Q,t)t^{N/2 -(N/2-s) -1}dt \cr
&\ \ + 
\int_1^\infty 
    \phi
   { \left\vert \left\vert \matrix{ -{\vec h} \cr
                     {\vec g} } \right\vert \right\vert }
   (Q^{-1},t)t^{N/2 -s -1}dt
\big\} \cr
& = { \exp -i2\pi {\vec g}\cdot{\vec h} \over \sqrt{det Q} }
 {\bf M}_{N/2 -s} 
    \phi
   { \left\vert \left\vert \matrix{ - {\vec h} \cr
                     {\vec g} } \right\vert \right\vert }
   (Q^{-1},\cdot)\ \ \ . & (A8) \cr
}
 $$
 Here, the change of variable $t \rightarrow 1/t$ has been made 
 to get the first term in the second line,  and $(A7)$ 
 has been substituted into the same term to get the next  line.
 Then, the same procedure has been repeated to get the last line. 
 Along with $(A3)$, this proves $(A5)$.
 
 We can see the pole-structure of the L.H.S. of $(A3)$, 
 i.e. of 
 $
 {\bf M}_s
 \phi
 { \left\vert \left\vert \matrix{ {\vec g} \cr
                     {\vec h} } \right\vert \right\vert }
 (Q,\cdot)
$
 in the third line of $(A8)$. Noting that 
 $\Gamma (s)$ has simple poles at $s=-k$ 
 $(k=0,1,2,\cdots)$ with residue ${ (-)^k \over  k! }$, 
 we see the above-mentioned pole- and  zero-structures of 
            $ Z
              {  \left\vert \left\vert \matrix{ {\vec g} \cr
                     {\vec h} } \right\vert \right\vert  }
  (Q,s)  $
   from $(A3)$.

\REF\one{See e.g., P.J.E. Peebles, {\it Principles of Physical Cosmology}, 
Princeton University Press, Princeton, 1993.}
\REF\two{ L.Z. Fang and H. Sato,  Comm. Theor. Phys. 
 {\bf 2}, 1055 (1983).}
\REF\three{A.A. Starobinsky,  JETP Lett. {\bf 57}, 622 (1993); 
      D. Stevens, D. Scott and J. Silk,  Phys. Rev. Lett.
        {\bf 71}, 20 (1993). } 
\REF\four{M. Seriu,  
Proceedings of the Indian Association for General Relativity
and Gravitation (1994);
 M. Seriu and T.P. Singh, Phys. Rev. {\bf D50}, 6165 (1994).}
\REF\five{J.A. Wheeler, Phys. Rev. {\bf 97}, 511 (1955).}
\REF\six{ C.W. Misner and J.A. Wheeler, 
 Ann. Phys. {\bf 2}, 525 (1957);
 J.A. Wheeler, {\sl Geometrodynamics }, 
Academic Press, New York,  1962; 
R. Sorkin, J. Phys. {\bf A10}, 717 (1977).}
\REF\seven{e.g. M.S. Morris, K.S. Thorne and U. Yurtsever, 
Phys. Rev. Lett. {\bf 61}, 1446 (1988).}
\REF\eight{e.g. G.W. Gibbons and J.B. Hartle,  Phys. Rev. {\bf D42},
 2458 (1990).}                              
\REF\nine{M. Visser,  Phys. Rev. {\bf D41}, 1116 (1990).}
\REF\ten{M. Seriu, Physics Letters {\bf B319}, 74 (1993);
  Vistas in Astronomy {\bf 37}, 637 (1993).}
\REF\eleven{B.A. Dubrovin, A.T. Fomenko and  S.P. Novikov, 
 {\it Modern Geometry - Methods and Applications } Part II, 
 Springer-Verlag, New York, 1985.  }
\REF\twelve{J.A.~Wheeler,  Ann. Phys.(N.Y.) {\bf 2}, 604 (1957). } 
\REF\thirteen{M. Kac, Am. Math. Mon. {\bf 73}(4), Part II, 1 (1966).} 
\REF\fourteen{J. Milnor, Proc. Nat. Acad. Sci. USA {\bf 51}, 542 (1964).}
\REF\fifteen{M.F. Vign\'eras, Ann. of Math. {\bf 112}, 21 (1980);
 A. Ikeda, J. Math. Soc. Japan  {\bf 35}, 437 (1983).}
\REF\sixteen{D. Deturck, H. Gluck, C. Gordon and D. Webb, in   
{\it Mechanics, Analysis and Geometry: 200 Years after Lagrange}, 
edited by M. Francaviglia,  
Elsevier Science Publishers B.V., Amsterdam, 1991. }
\REF\seventeen{T. Sunada, {\it Fundamental Groups and the Laplacian}, 
Kinokuniya Publishing, Tokyo,  1988.}
\REF\eighteen{B.S. DeWitt, Phys. Rev. {\bf 160}, 1113 (1967). }
\REF\nineteen{P. Epstein, Math. Ann. {\bf 56}, 615 (1903); 
{\it ibid} {\bf 63}, 205 (1907).}
\REF\twenty{C. L. Siegel, {\it Advanced Analytic Number Theory}, 
  Tata Institute of Fundamental Research, Bombay,  1980.}
\REF\twentyone{L.D. Landau and E.M. Lifshitz, {\sl Quantum Mechanics 
(Non-relativistic Theory)} 3rd Ed., \S 41 and \S 53, Pergamon Press,
 Oxford,  1977.}  
\REF\twentytwo{H.D. Zeh, Phys. Lett. {\bf A}, 9 (1986).}
\REF\twentythree{C. Kiefer, Class. Quantum Grav. {\bf 4}, 1369 (1987).}
\REF\twentyfour{T. Padmanabhan, Phys. Rev. {\bf D39 }, 2924  (1989).}
\REF\twentyfive{J.J. Halliwell and S.W. Hawking, 
Phys. Rev. {\bf D 31}, 1777 (1985).}
\REF\twentysix{I. Chavel, {\it Eigenvalues in Riemannian Geometry}, 
 Academic Press, Orland,  1984.  }
 \REF\twentyseven{E.B. Davies, {\it Heat Kernels and Spectral Theory}, 
 Cambridge University Press, Cambridge, 1989.}
 \REF\twentyeight{J. Wolf, {\it Spaces of Constant Curvature}, 
 McGraw-Hill, New York, 1967.}
\REF\twentynine{B. Hatfield, {\sl Quantum Field Theory of 
Point Particles and Strings}, Addison -Wesley, Redwood City,
California,  1992.}
 \REF\thirty{J.S. Dowker, J. Math. Phys. {\bf 28}, 33 (1987); 
 Phys. Rev. {\bf D40}, 1938 (1989). }
 \REF\thirtyone{N.L. Balaza and A.  Voros, Phys. Rep. {\bf 143}, 109 
 (1986). }
 \REF\thirtytwo{M. Nakahara, {\sl Geometry, Topology and Physics},  
 IOP Publishing, Bristol,  1990.}
 \REF\thirtythree{L.V. Ahlfors, {\it Complex Analysis}, McGraw-Hill,
 New York,  1979.}
 \REF\thirtyfour{P. Cartier, in 
  {\it From Number Theory to Physics}, 
 edited by  M. Waldschmidt, P. Moussa, J.-M. Luck and 
 C. Itzykson,  Springer-Verlag, Berlin, 
  1992.}

\refout

\endpage

\centerline{Figure Captions}

\item{Figure\  1-a}
 The spectral diagram for $T^2$ ($\tau^1 \ne 0$). 
 Symbols `$\bf i$' and `$\bf ii$' 
 represent the multiplicity 1 and 2, respectively. 
 For instance, the multiplicity of the eigenvalue $\lambda_{(0,0)}$ 
 (zero-mode) is 1 and the same of the eigenvalue $\lambda_{(2,3)}$ is 2.   
\item{Figure\  1-b}
The spectral diagram for $T^2$ ($\tau^1 = 0$). The symbol `$\bf iv$'
  represents multiplicity 4.
\item{Figure\  2}
Klein's bottle constructed by ${\bf R}^2 / G$. All line segments 
 which are parallel to each other  
should be identified respecting the direction shown by 
an  arrow sign.
A special  letter  is drawn to visualize the way of identification.
\item{Figure\  3}
The spectral diagram for  Klein's bottle ${\bf R}P^2 \# {\bf R}P^2$.
\item{Figure\  4}
A $\lambda_k -k$ plot for a torus with $(\tau^1, \tau^2)=(0.1, 1)$.
\item{Figure\  5}
A  $4\pi N(\Lambda) - \Lambda$ 
plot for a torus with $(\tau^1, \tau^2)=(0.1, 1)$.
\item{Figure\  6-a}
Two tori with $(\tau^1, \tau^2)=(0, 1)$ and $(\tau^1, \tau^2)=(0, 2)$.
\item{Figure\  6-b}
A $d_\Lambda - \Lambda$ plot for the tori shown in $Figure$ $6-a$.
\item{Figure\  7-a}
Two Klein's bottles  with 
$(\tau^1, \tau^2)=(0, 10)$ and $(\tau^1, \tau^2)=(0, 100)$.
\item{Figure\  7-b}
A $d_\Lambda - \Lambda$ plot for Klein's bottles
 shown in $Figure$ $7-a$.
\item{Figure\  8-a}
A torus with  $(\tau^1, \tau^2)=(0, 1)$ (a regular square) 
and Klein's bottle with 
 $(\tau^1, \tau^2) = (0, 10)$ (a rectangle).
\item{Figure\  8-b}
A $d_\Lambda - \Lambda$ plot for a torus and Klein's bottle
 shown in $Figure$ $8-a$.
 \item{Figure\  9-a}
Spectral distances between two tori.   
Parameters $(\tau^1, \tau^2)$ for two tori and the spectral distance 
between them are indicated.
\item{Figure\  9-b}
Spectral distances between  two Klein's bottles.   
Parameters $(\tau^1, \tau^2)$ for two Klein's bottles 
and the spectral distance 
between them are indicated.
\item{Figure\  9-c}
Spectral distances between a torus and  Klein's bottle.
From top to bottom, 
parameters  $(\tau^1, \tau^2)$ for a torus and Klein's bottle, and 
the spectral distance between them are indicated, in this order.
\item{Figure\  10}
Distances between two tori defined by the DeWitt metric.
\item{Figure\  11}
The spectral distance between a 2-sphere and a real projective space 
with  identical 2-volumes (area).
\item{Figure\  12}
The spectral distance between two 2-spheres  with different 2-volumes, 
 1.0 and  1.1.

\end